\documentclass[twocolumn]{aastex63}
\hypersetup{linkcolor=blue,urlcolor=blue}

\shorttitle{ALMA and NOEMA [\ion{C}{2}] Survey of Reionization-Era Quasars}
\shortauthors{Wang et al.}
\begin{document}

\title{A Spatially Resolved [\ion{C}{2}] Survey of 31 $z\sim7$ Massive Galaxies Hosting Luminous Quasars}

\correspondingauthor{Feige Wang}
\email{feigewang@email.arizona.edu}

\author[0000-0002-7633-431X]{Feige Wang}
\affil{Steward Observatory, University of Arizona, 933 North Cherry Avenue, Tucson, AZ 85721, USA}

\author[0000-0001-5287-4242]{Jinyi Yang}
\thanks{Strittmatter Fellow}
\affil{Steward Observatory, University of Arizona, 933 North Cherry Avenue, Tucson, AZ 85721, USA}

\author[0000-0003-3310-0131]{Xiaohui Fan}
\affil{Steward Observatory, University of Arizona, 933 North Cherry Avenue, Tucson, AZ 85721, USA}

\author[0000-0001-9024-8322]{Bram Venemans}
\affil{Max Planck Institut f\"ur Astronomie, K\"onigstuhl 17, D-69117, Heidelberg, Germany}
\affil{Leiden Observatory, Leiden University, P.O. Box 9513, NL-2300 RA Leiden, The Netherlands}

\author[0000-0002-2662-8803]{Roberto Decarli}
\affil{INAF--Osservatorio di Astrofisica e Scienza dello Spazio, via Gobetti 93/3, I-40129, Bologna, Italy}

\author[0000-0002-2931-7824]{Eduardo Ba\~nados}
\affil{Max Planck Institut f\"ur Astronomie, K\"onigstuhl 17, D-69117, Heidelberg, Germany}

\author[0000-0003-4793-7880]{Fabian Walter}
\affil{Max Planck Institut f\"ur Astronomie, K\"onigstuhl 17, D-69117, Heidelberg, Germany}

\author[0000-0002-3026-0562]{Aaron J. Barth}
\affil{Department of Physics and Astronomy, 4129 Frederick Reines Hall, University of California, Irvine, CA, 92697-4575, USA}

\author[0000-0002-1620-0897]{Fuyan Bian}
\affil{European Southern Observatory, Alonso de C\'ordova 3107, Casilla 19001, Vitacura, Santiago 19, Chile}

\author[0000-0003-0821-3644]{Frederick B. Davies}
\affil{Lawrence Berkeley National Laboratory, CA 94720-8139, USA}
\affil{Max Planck Institut f\"ur Astronomie, K\"onigstuhl 17, D-69117, Heidelberg, Germany}

\author[0000-0003-2895-6218]{Anna-Christina Eilers}
\affil{MIT Kavli Institute for Astrophysics and Space Research, 77 Massachusetts Avenue, Cambridge, Massachusetts 02139, USA}

\author[0000-0002-6822-2254]{Emanuele Paolo Farina}
\affiliation{Gemini Observatory, NSF’s NOIRLab, 670 N A’ohoku Place, Hilo, Hawai'i 96720, USA}

\author[0000-0002-7054-4332]{Joseph F. Hennawi}
\affil{Leiden Observatory, Leiden University, P.O. Box 9513, NL-2300 RA Leiden, The Netherlands}
\affil{Department of Physics, University of California, Santa Barbara, CA 93106-9530, USA}

\author[0000-0001-6239-3821]{Jiang-Tao Li}
\affil{Department of Astronomy, University of Michigan, 311 West Hall, 1085 S. University Ave, Ann Arbor, MI, 48109-1107, USA}

\author[0000-0002-5941-5214]{Chiara Mazzucchelli}
\affil{Instituto de Estudios Astrof\'{\i}sicos, Facultad de Ingenier\'{\i}a y Ciencias, Universidad Diego Portales, Avenida Ejercito Libertador 441, Santiago, Chile.}

\author{Ran Wang}
\affil{Kavli Institute for Astronomy and Astrophysics, Peking University, Beijing 100871, China}

\author[0000-0002-7350-6913]{Xue-Bing Wu}
\affil{Kavli Institute for Astronomy and Astrophysics, Peking University, Beijing 100871, China}
\affil{Department of Astronomy, School of Physics, Peking University, Beijing 100871, China}

\author[0000-0002-5367-8021]{Minghao Yue}
\affil{MIT Kavli Institute for Astrophysics and Space Research, 77 Massachusetts Avenue, Cambridge, Massachusetts 02139, USA}

\begin{abstract}
The [\ion{C}{2}] 158 $\mu$m emission line and the underlying far-infrared (FIR) dust continuum are important tracers for studying star formation and kinematic properties of early galaxies. We present a survey of the [\ion{C}{2}] emission lines and FIR continua of 31 luminous quasars at $z>6.5$ using the Atacama Large Millimeter Array (ALMA) and the NOrthern Extended Millimeter Array (NOEMA) at sub-arcsec resolution. This survey more than doubles the number of quasars with [\ion{C}{2}] and FIR observations at these redshifts and enables statistical studies of quasar host galaxies deep into the epoch of reionization. We detect [\ion{C}{2}] emission in 27 quasar hosts with a luminosity range of $L_{\rm [CII]}=(0.3-5.5)\times10^9~L_\odot$ and detect the FIR continuum of 28 quasar hosts with a luminosity range of $L_{\rm FIR}=(0.5-13.0)\times10^{12}~L_\odot$. Both $L_{\rm [CII]}$ and $L_{\rm FIR}$ are correlated ($\rho\simeq0.4$) with the quasar bolometric luminosity, albeit with substantial scatter. The quasar hosts detected by ALMA are clearly resolved with a median diameter of $\sim$5 kpc. About $40$\% of the quasar host galaxies show a velocity gradient in [\ion{C}{2}] emission, while the rest show either dispersion-dominated or disturbed kinematics. 
Basic estimates of the dynamical masses of the rotation-dominated host galaxies yield $M_{\rm dyn}=(0.1-7.5)\times10^{11}~M_\odot$. 
Considering our findings alongside those of literature studies, we found that the ratio between $M_{\rm BH}$ and $M_{\rm dyn}$ is about ten times higher than that of local $M_{\rm BH}-M_{\rm dyn}$ relation on average but with substantial scatter (the ratio difference ranging from $\sim$0.6 to 60) and large uncertainties.
\end{abstract}


\section{Introduction} \label{sec_intro}
The discovery of the strong correlations between the mass of supermassive black holes (SMBHs) and the physical properties (e.g., stellar velocity dispersion, bulge mass, dynamical mass) of their host galaxies (i.e., the $M-\sigma_{\ast}$ relation) in the local Universe indicates that SMBHs co-evolve with their host galaxies \citep[see][ and references therein]{KH13}. Such relations are thought to arise from the fact that the energy released by accreting SMBHs is able to affect gas kinematic and star formation activity in the host galaxy, thus regulating both the dynamical properties and the stellar mass of the host \citep[e.g.][]{DiMatteo05,Hopkins08}. Therefore, investigating the SMBH-galaxy relations at high redshift is crucial for understanding galaxy formation and evolution at early cosmic times. However, luminous quasars in the rest-frame UV significantly outshine their hosts, prohibiting the detection of the stellar light or the measurement of the $M-\sigma_{\ast}$ relation in luminous $z\gtrsim6$ SMBH-galaxy systems even with the {Hubble Space Telescope} \citep{Mechtley12,Marshall20}.

Thanks to the superb sensitivity of the {Atacama Large Millimeter/submillimeter Array} (ALMA) and the {NOrthern Extended Millimeter Array} (NOEMA), the host galaxies of more than 60 $z\gtrsim6$ quasars have been detected in rest-frame far-infrared dust continuum and the fine structure [\ion{C}{2}] 158 $\rm\mu m$ line \citep[e.g.,][]{Walter09,Wang13,Wang16,Willott15,Venemans16,Decarli18,Feruglio18,Venemans18,Izumi19,Izumi21,Wang19b,Yang19b,Eilers20,Pensabene20}. Several key findings have been established based on these observations. 
Firstly, copious amounts of dust and gas are present in the majority of these $z > 6$ quasar host galaxies \citep[see][for reviews]{Carilli13, Fan23}.
These galaxies, significantly enriched with metals and powered by prodigious star formation with star formation rate (SFR) of $\rm \gtrsim100 ~M_\odot~yr^{-1}$,  are significantly more massive than typical star-forming galaxies found in deep fields \citep[e.g.,][]{Bouwens21}. 
Secondly, kinematic modeling of several high spatial resolution (sub-kpc scale) observations of the [\ion{C}{2}] line suggests that the most luminous systems seem to have over-massive SMBHs compared with expectations from the local $M-\sigma_{\ast}$ relation while fainter systems still follow the local $M-\sigma_{\ast}$ relation \citep[e.g.,][]{Banados19,Wang19b,Venemans19,Pensabene20,Neeleman21, Farina22}, suggesting that we are witnessing the establishment of the $M-\sigma_{\ast}$ relation in the early Universe.
Thirdly, the existence of dust-rich companion galaxies adjacent to some fraction of $z\gtrsim6$ quasars indicates that the earliest SMBHs grew in galaxy-rich environment, and that major mergers may be important drivers for rapid SMBH and host galaxy growth \citep[e.g.,][]{Decarli17,Mazzucchelli19,Venemans20, Meyer22}.

The majority of the discoveries mentioned above were based on observations of $z\sim6$ quasars, while observations of higher redshift (e.g., $z>6.5$) systems are still limited to only a dozen \citep[e.g.,][]{Venemans16,Neeleman21,Izumi21}. 
On the other hand, recent studies show that the spatial density of quasars rises extraordinarily rapidly (by a factor of six) from $z\sim7$ to $z\sim6$, significantly faster than that from $z\sim6$ to $z\sim3$ \citep{Wang19a}, indicating a rapid buildup of the earliest SMBHs in this short timescale. 
In the last few years, the number of known $z>6.5$ quasars has dramatically increased \citep[e.g.,][]{Wang17,Wang18,Wang19a,Wang21a,Banados18,Matsuoka18,Matsuoka19b,Reed19,Yang19a,Yang20a, Yang20b,Yang21}. This motivated us to perform a comprehensive [\ion{C}{2}] survey of quasar host galaxies at $z\sim7$. The main science goals of this survey are 
to characterize star formation activities in the host galaxies of the earliest SMBHs, 
to constrain the dynamical masses of quasar host galaxies and thus investigate the $M_{\rm BH}-M_{\rm dynamical}$ relation at $z\sim7$, 
to perform a census of [\ion{C}{2}] emitters in quasar vicinities, 
and to probe cosmic reionization history and black hole growth by combining the [\ion{C}{2}] redshift measurements with optical and infrared spectroscopy. 

In this paper, we present an overview of our spatially resolved [\ion{C}{2}] survey of 31 luminous $z\sim7$ quasars and discuss the properties of quasar host galaxies based on [\ion{C}{2}] and dust emission. 
In \S \ref{sec:obs}, we will describe the quasar sample construction, ALMA and NOEMA observations, and data reduction. 
In \S \ref{sec:rlt}, we will present the [\ion{C}{2}] line properties, continuum luminosity, SFR, as well as the size measurements of quasar host galaxies.  
In \S \ref{sec:mass}, we will highlight the diverse kinematic properties of quasar host galaxies, present first-order constraints on their dynamical masses, and discuss the $M-\sigma_{\ast}$ relation of these systems. 
Finally, we will conclude in \S \ref{sec:summary}.
Throughout the paper, we adapt a flat cosmology model with $H_0=70.0~{\rm km~s^{-1}~Mpc^{-1}}$, $\rm \Omega_M=0.3$, and $\Omega_\Lambda=0.7$.

\begin{figure}
\centering
\includegraphics[width=0.48\textwidth]{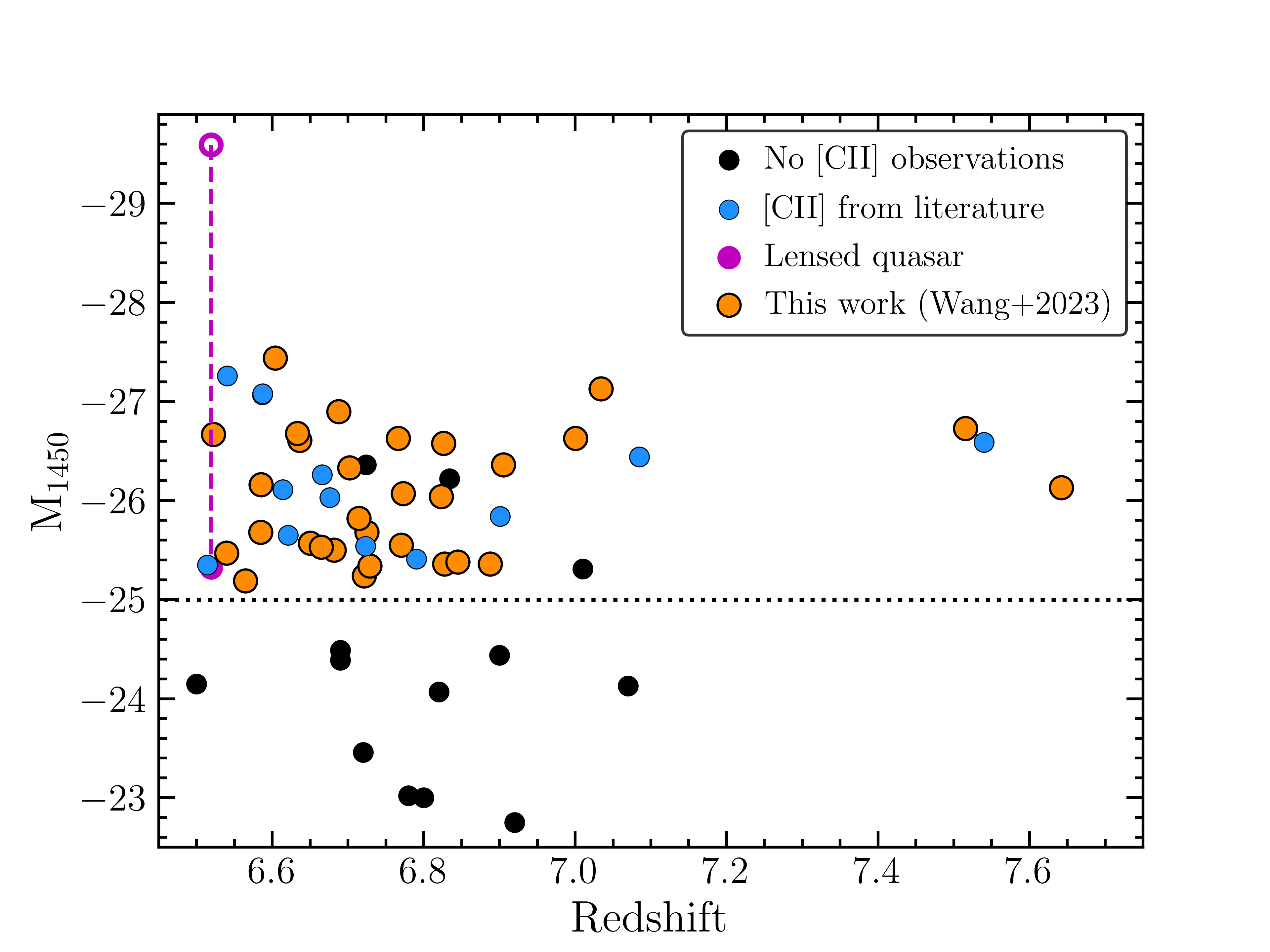}
\caption{The redshift and $M_{1450}$ distribution of quasars known at $z\ge6.5$. This [\ion{C}{2}] survey aims to observe all $z\ge6.5$ quasars with $M_{1450}<-25.0$ (black dotted line).
The quasars J0244$-$5008 ($z=6.724$) and J0020$-$3653 ($z=6.834$) were proposed in ALMA Cycle 5 by the discovery team \citep{Reed19} and were excluded from this survey. Another quasar J2356+0017 at $z=7.01$ was discovered \citep{Matsuoka19b} after our survey begins. The ALMA observation of the gravitationally lensed quasar, J0439+1634, is also reported in this work.
\label{fig:m1450}}
\end{figure}

\begin{deluxetable*}{cccccccccrcr}
\tablecaption{The basic information and observing log of 31 $z>6.5$ quasars studied here.}\label{tbl:basic}
\setlength{\tabcolsep}{2pt}
\tabletypesize{\scriptsize}
\tablehead{\colhead{Name} & \colhead{RA} & \colhead{DEC}& \colhead{$z_{\rm MgII}$}& \colhead{$M_{\rm 1450}$} & \colhead{Obs.} & \colhead{Program ID} & \colhead{Exposure} & \colhead{Beam} & \colhead{$\rm PA_{Beam}$} & \colhead{RMS (30 km/s)} & \colhead{Ref.\tablenotemark{b}}}
\startdata
                        &      J2000        &       J2000          &             &             &             &                          & [min] & [$^{\prime\prime}\times^{\prime\prime}$]     & [deg]       & [mJy/beam]      & \\
(1)          &   (2)       &     (3)       &    (4)    &    (5)    &    (6)    &    (7)    &    (8)    &    (9)    &    (10)    &    (11)    &    (12)   \\
\hline
J0038$-$1527 & 00:38:36.10 & $-$15:27:23.6 & 6.999$\pm$0.001 & $-$27.13 & ALMA & 2018.1.01188.S & 12.6 & $0.60\times0.54$ & $-$88.06 & 0.38 & 1/2 \\
J0213$-$0626 & 02:13:16.94 & $-$06:26:15.2 & 6.72\tablenotemark{a} & $-$25.24 & ALMA & 2018.1.01188.S & 12.6 & $0.61\times0.52$ & 83.80 & 0.32 & 3/3 \\
J0218$+$0007 & 02:18:47.04 & $+$00:07:15.2 & 6.766$\pm$0.004 & $-$25.55 & ALMA & 2018.1.01188.S & 26.2 & $0.38\times0.36$ & 70.84 & 0.3 & 4/2 \\
J0224$-$4711 & 02:24:26.54 & $-$47:11:29.4 & 6.527$\pm$0.001 & $-$26.67 & ALMA & 2018.1.01188.S & 13.61 & $0.57\times0.56$ & 36.54 & 0.35 &5/6 \\
J0229$-$0808 & 02:29:35.25 & $-$08:08:23.0 & 6.727 & $-$25.68 & ALMA & 2018.1.01188.S & 12.6 & $0.39\times0.34$ & 53.03 & 0.38 & 7/7 \\
J0246$-$5219 & 02:46:55.90 & $-$52:19:49.9 & 6.857$\pm$0.019 & $-$25.36 & ALMA & 2018.1.01188.S & 40.32 & $0.38\times0.35$ & 10.58 & 0.27 & 8/2 \\
J0252$-$0503 & 02:52:16.64 & $-$05:03:31.8 & 6.990$\pm$0.017 & $-$26.63 & ALMA & 2018.1.01188.S & 13.1 & $0.49\times0.38$ & 65.91 & 0.4 & 8/2 \\
                        &                    &                         &                    &                          & NOEMA & W18EJ & 150  & $1.76\times0.77$ & 27.24 & 0.82 &  \\ 
J0313$-$1806 & 03:13:43.84 & $-$18:06:36.4 & 7.611$\pm$0.004 & $-$26.13 & ALMA & 2019.A.00017.S & 57.46 & $0.60\times0.43$ & $-$74.53 & 0.29 & 9/2 \\
J0319$-$1008 & 03:19:41.66 & $-$10:08:46.0 & 6.816$\pm$0.004 & $-$25.36 & ALMA & 2019.1.01025.S & 13.1 & $0.39\times0.33$ & 53.49 & 0.44 & 8/2 \\
J0411$-$0907 & 04:11:28.63 & $-$09:07:49.7 & 6.827$\pm$0.006 & $-$26.58 & ALMA & 2018.1.01188.S & 13.1 & $0.69\times0.55$ & 74.83 & 0.37 & 10/2\\
J0430$-$1445 & 04:30:43.66 & $-$14:45:41.2 & 6.718 & $-$25.82 & ALMA & 2019.1.01025.S & 12.6 & $0.37\times0.33$ & $-$75.80 & 0.43 & 7/7 \\
J0439$+$1634 & 04:39:47.10 & $+$16:34:15.8 & 6.519$\pm$0.003 & $-$29.59 & ALMA & 2018.1.01188.S & 15.12 & $0.87\times0.49$ & $-$47.07 & 0.5 & 11/2 \\
J0525$-$2406 & 05:25:59.68 & $-$24:06:23.0 & 6.543$\pm$0.002 & $-$25.47 & ALMA & 2019.1.01025.S & 12.6 & $0.37\times0.33$ & $-$76.26 & 0.36 & 4/2 \\
J0706$+$2921 & 07:06:26.38 & $+$29:21:05.5 & 6.593$\pm$0.000 & $-$27.44 & ALMA & 2018.1.01188.S & 28.22 & $0.90\times0.63$ & $-$11.47 & 0.4 & 10/2 \\
J0829$+$4117 & 08:29:31.98 & $+$41:17:40.9 & 6.773$\pm$0.007 & $-$26.07 & NOEMA & W18EJ &  150 & $0.99\times0.59$ & 8.23 & 0.76 & 10/2 \\ 
J0837$+$4929 & 08:37:37.83 & $+$49:29:00.6 & 6.702$\pm$0.001 & $-$26.33 & NOEMA & W18EJ & 138 & $1.15\times0.61$ & $-$19.47 & 0.67 & 10/2 \\ 
J0839$+$3900 & 08:39:46.88 & $+$39:00:11.4 & 6.905$\pm$0.000 & $-$26.36 & NOEMA & W18EJ & 150 & $1.00\times0.57$ & 12.29 & 0.74 & 10/2 \\ 
J0910$+$1656 & 09:10:13.65 & $+$16:56:30.2 & 6.719$\pm$0.005 & $-$25.34 & ALMA & 2018.1.01188.S & 15.12 & $0.68\times0.56$ & $-$4.87 & 0.31 & 10/2 \\
J0910$-$0414 & 09:10:54.54 & $-$04:14:06.9  & 6.610$\pm$0.003 & $-$26.61 & ALMA & 2018.1.01188.S & 19.15 & $0.62\times0.55$ & 69.15 & 0.27 & 10/2 \\
J0921$+$0007 & 09:21:20.56 & $+$00:07:22.9 & 6.565$\pm$0.000 & $-$25.19 & ALMA & 2018.1.01188.S & 13.1 & $0.60\times0.55$ & 58.55 & 0.31 & 4/2 \\
J0923$+$0402 & 09:23:47.12 & $+$04:02:54.6 & 6.612$\pm$0.002 & $-$26.68 & ALMA & 2018.1.01188.S & 13.61 & $0.61\times0.57$ & 38.77 & 0.33 & 10/2\\
J0923$+$0753 & 09:23:58.99 & $+$07:53:48.7 & 6.682$\pm$0.002 & $-$25.50 & ALMA & 2019.1.01025.S & 29.24 & $0.40\times0.34$ & 37.87 & 0.26 & 4/2 \\
J1007$+$2115 & 10:07:58.26 & $+$21:15:29.2 & 7.476$\pm$0.013 & $-$26.73 & ALMA & 2019.1.01025.S & 15.12 & $0.48\times0.36$ & $-$4.71 & 0.41 & 12/2 \\
J1058$+$2930 & 10:58:07.72 & $+$29:30:41.7 & 6.585$\pm$0.005 & $-$25.68 & ALMA & 2019.1.01025.S & 14.62 & $0.53\times0.35$ & 17.00 & 0.59 & 4/2 \\
J1104$+$2134 & 11:04:21.58 & $+$21:34:28.9 & 6.766$\pm$0.005 & $-$26.63 & ALMA & 2018.1.01188.S & 16.13 & $0.70\times0.59$ & 8.46 & 0.37 & 10/2 \\
J1129$+$1846 & 11:29:25.37 & $+$18:46:24.3 & 6.824$\pm$0.001 & $-$26.04 & ALMA & 2019.1.01025.S & 16.13 & $0.43\times0.34$ & 8.86 & 0.35 & 13/2 \\
J1135$+$5011 & 11:35:08.92 & $+$50:11:32.6 & 6.579$\pm$0.001 & $-$26.16 & NOEMA & W18EJ & 177 & $0.90\times0.77$ & $-$40.45 & 0.58 & 10/2 \\ 
J1216$+$4519 & 12:16:27.58 & $+$45:19:10.7 & 6.648$\pm$0.003 & $-$25.57 & NOEMA & W18EJ & 148 & $1.21\times0.66$ & 96.55 & 0.93 & 10/2 \\ 
J2002$-$3013 & 20:02:41.59 & $-$30:13:21.7 & 6.673$\pm$0.001 & $-$26.90 & ALMA & 2019.1.01025.S & 12.6 & $0.38\times0.33$ & 88.60 & 0.47 & 4/2 \\
J2102$-$1458 & 21:02:19.23 & $-$14:58:53.9 & 6.652$\pm$0.003 & $-$25.53 & ALMA & 2018.1.01188.S & 13.1 & $0.59\times0.50$ & $-$67.68 & 0.37 & 10/2 \\
J2211$-$6320 & 22:11:00.60 & $-$63:20:55.9 & 6.832$\pm$0.015 & $-$25.38 & ALMA & 2019.1.01025.S & 32.26 & $0.54\times0.45$ & $-$18.86 & 0.36 &8/2
\enddata
{{\bf Notes:} (1) Quasar name, (2-3) R.A. and decl. (J2000), (4) redshift from \ion{Mg}{2} line, (5) absolute magnitude at rest-frame 1450 \AA, (6) observatory, (7) program ID, (8) on-source exposure time, (9) beam size, (10) position angle of the beam, (11) rms noise per 30 km/s bin, (12) reference for quasar discovery, $z_{\rm MgII}$ and $M_{1450}$ measurements, see below. }
\tablenotetext{a}{There is no infrared spectroscopic observation for this object; the redshift was measured from the Ly$\alpha$ line by \cite{Matsuoka18}.}
\tablenotetext{b}{References for the quasar discovery (first number) and for $z_{\rm MgII}$ and $M_{\rm 1450}$ measurements (second number).
{\bf References:} 1: \cite{Wang18}; 2: \cite{Yang21}; 3: \cite{Matsuoka18}; 4: Yang et al. in prep; 5:  \cite{Reed17}; 6:  \cite{Wang21b}; 7: Ba\~nados et al. in prep; 
8: \cite{Yang19a}; 9: \cite{Wang21a}; 10: \cite{Wang19a}; 11: \cite{Fan19}; 12: \cite{Yang20a}; 13: \cite{Banados21}.
}
\end{deluxetable*}

\section{Observations and Data Reduction}\label{sec:obs}
The parent sample of our [\ion{C}{2}] survey was built to include all 56 $z>6.5$ quasars with $M_{1450}<-25.0$ known at the time of designing this program. The redshift and $M_{1450}$ distributions of these quasars and all other known quasars at $z>6.5$ are shown in Figure \ref{fig:m1450}. In the literature, 13 quasars had either ALMA or NOEMA observations at the time of designing this program and thus were not targeted by this survey. An additional two quasars, J0244$-$5008 and J0020$-$3653 \citep{Reed19}, were proposed to ALMA in Cycle 5 by the quasar discovery team and were also excluded from this survey. 
We observed the remaining 31 $z>6.5$ quasars with $M_{1450}<-25.0$ with ALMA and NOEMA: 26 quasars with $\rm Decl. < 30^\circ$ were targeted with ALMA and the other 5 quasars with $\rm Decl.>30^\circ$ were observed with NOEMA. One quasar in our sample, J0252--0503, was observed by both ALMA and NOEMA. 
The NOEMA observation of the gravitationally lensed quasar J0439+1634 at $z=6.52$ was published in \cite{Yang19b} and the high-resolution (C43-5) ALMA observation of this object was published in \cite{Yue21}. In this paper, we present the ALMA C43-3 observation of J0439+1634 obtained from this program. 
Although the ALMA observations of J1007+2115 and J0313--1806 were reported in \cite{Yang20a} and \cite{Wang21a}, respectively, we include the observations of these two quasars in this paper for completeness. 
The basic information of these 31 quasars as well as the observation details are listed in Table \ref{tbl:basic}.
In the following sections, we only reports the measurements for objects targeted by this survey but includes objects from literature for statistically analyses.

\subsection{ALMA observation and data reduction}
Our ALMA observations span two cycles with one Cycle 6 program (program ID: 2018.1.01188.S, PI: F. Wang) and two Cycle 7 programs (program IDs: 2019.1.01025.S and 2019.A.00017.S, PI: F. Wang). These observations were designed to reach a flux sensitivity of $\sim$0.3 mJy/beam per 100 $\rm km ~s^{-1}$ and a resolution of 0\farcs4--0\farcs7, in order to detect the  [\ion{C}{2}] emission of nearly all quasars and marginally resolve the most extended quasar host galaxies. We tuned two spectral windows (SPW, 0 and 1) centered at the expected frequency of [\ion{C}{2}] and the other two SPWs (2 and 3) centered at about 15 GHz away from the expected [\ion{C}{2}] line for observing the continuum emission. Observations were carried out between 2018 December and 2020 March with 42--50 12m antennas. The on-source exposure time for each target was designed to be $\sim15$ minutes and targets were observed with either the C43-3 or C43-4 configuration, with the final on-source time depends on the observing conditions. The detailed ALMA observational information is listed in Table \ref{tbl:basic}. 

All the ALMA data were processed using the CASA \citep{CASA} pipeline for ALMA using the default calibration procedure. The Cycle 6 data were calibrated with CASA version 5.4.0 and the Cycle 7 data were calibrated with CASA version 5.6.1. All the data were then analyzed using CASA version 5.6.1. We imaged the data cubes using Briggs cleaning via the CASA task {\tt tclean} with robustness parameter $r=2.0$, corresponding to natural weighting, to maximize the signal-to-noise ratio (S/N) of our observations. Since we will compare our results with those of lower redshift  ($z\sim6$) quasars from \cite{Decarli18}, the imaging process was designed to closely follow that used by \cite{Decarli18}. The imaging process includes the following steps:

\begin{figure*}
\centering
\includegraphics[width=0.98\textwidth]{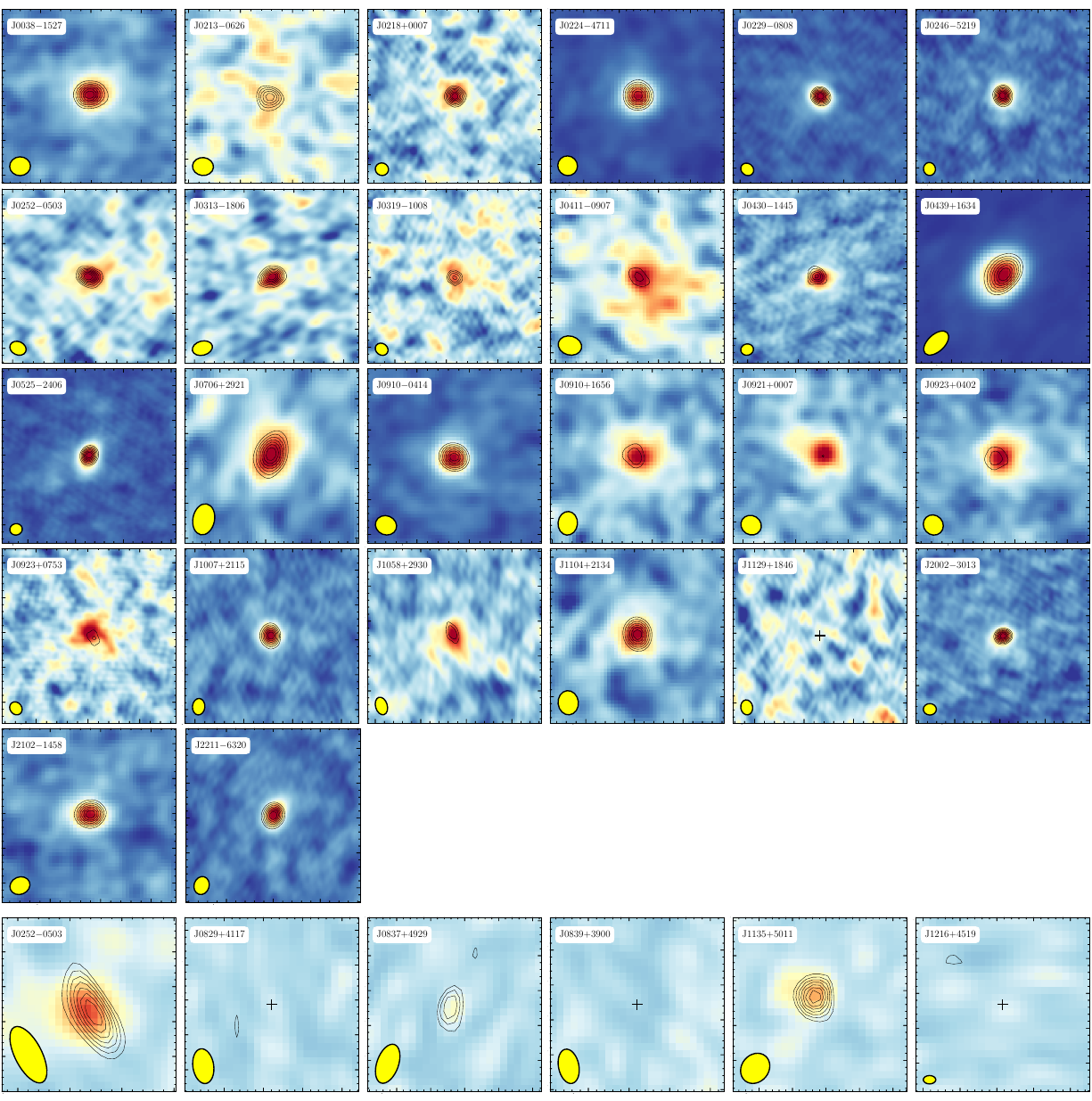}
\caption{ALMA (top 26 panels) and NOEMA (bottom 6 panels) thumbnails  of $z>6.5$ quasars showing the continuum-subtracted [\ion{C}{2}] line maps in color and the continuum in contours. Each panel is $5^{\prime\prime} \times 5^{\prime\prime}$ wide. North is up and east to the left. The solid black crosses on the [\ion{C}{2}] undetected objects mark the expected emission position. The solid black contours mark the +3, 4, 5, 6, 7, 8, 9 isophotes for those objects with continuum peak detected at $<10\sigma$, or 30\%, 40\%, 50\%, 60\%, 70\%, 80\%, 90\% of the continuum peak emission for those objects with $\ge10\sigma$  peak detections.
The synthesized beam of the observations is shown in the bottom-left corner of each panel. 
\label{fig:map1}}
\end{figure*}

(1) We generate a dirty continuum + line map for each target by collapsing the data in SPWs 0\&1 with {\tt tclean}. This map is used for identifying the positions of detected sources, creating box masks for detected sources and estimating the sensitivity of the observations.

(2) We then create data cubes for SPWs 0\&1 and 2\&3 with {\tt tclean}. During this step, we use the object mask regions generated from last step and re-bin the data cube to $\rm 23.4~MHz$ ($\rm\sim 30~ km~s^{-1}$ at the quasar redshifts).

(3) We extract a 1D spectrum of each object with an aperture diameter of 1\farcs5. Note that \cite{Decarli18} extracted 1D spectra on a single-pixel basis because their data have a lower resolution. 

(4) The spectra extracted from the SPWs 0\&1 data cubes are then fitted with a flat continuum and a Gaussian function. The continuum flux, line flux, line peak and line width are listed in Table \ref{tbl:c2fit}.

(5) We then subtract the continuum for the data cube from SPWs 0\&1 using the {\tt uvcontsub} task in the UV plane. 

(6) The [\ion{C}{2}] line map is collapsed from the data cube in the frequency range set by the line peak $\pm1.4\sigma_{\rm line}$ from the 1D spectral fitting generated in Step 4. 
For a Gaussian line profile, this integration corresponds to recovering $\sim$83\% of the total line flux. 
For the object (J1129+1846) without a successful Gaussian fitting from the 1D spectra, we assume the line has the same redshift as that derived from the \ion{Mg}{2} line and a width of FWHM = $\rm 300 ~km~s^{-1}$.  

(7) The continuum maps are generated by collapsing the entire line-free channels in SPWs 0\&1\&2\&3.

The [\ion{C}{2}] line maps and the continuum contours are shown in Figure \ref {fig:map1} and the extracted 1D spectra are shown in Figure \ref{fig:spectra1}.

\begin{figure*}
\centering
\includegraphics[width=0.98\textwidth]{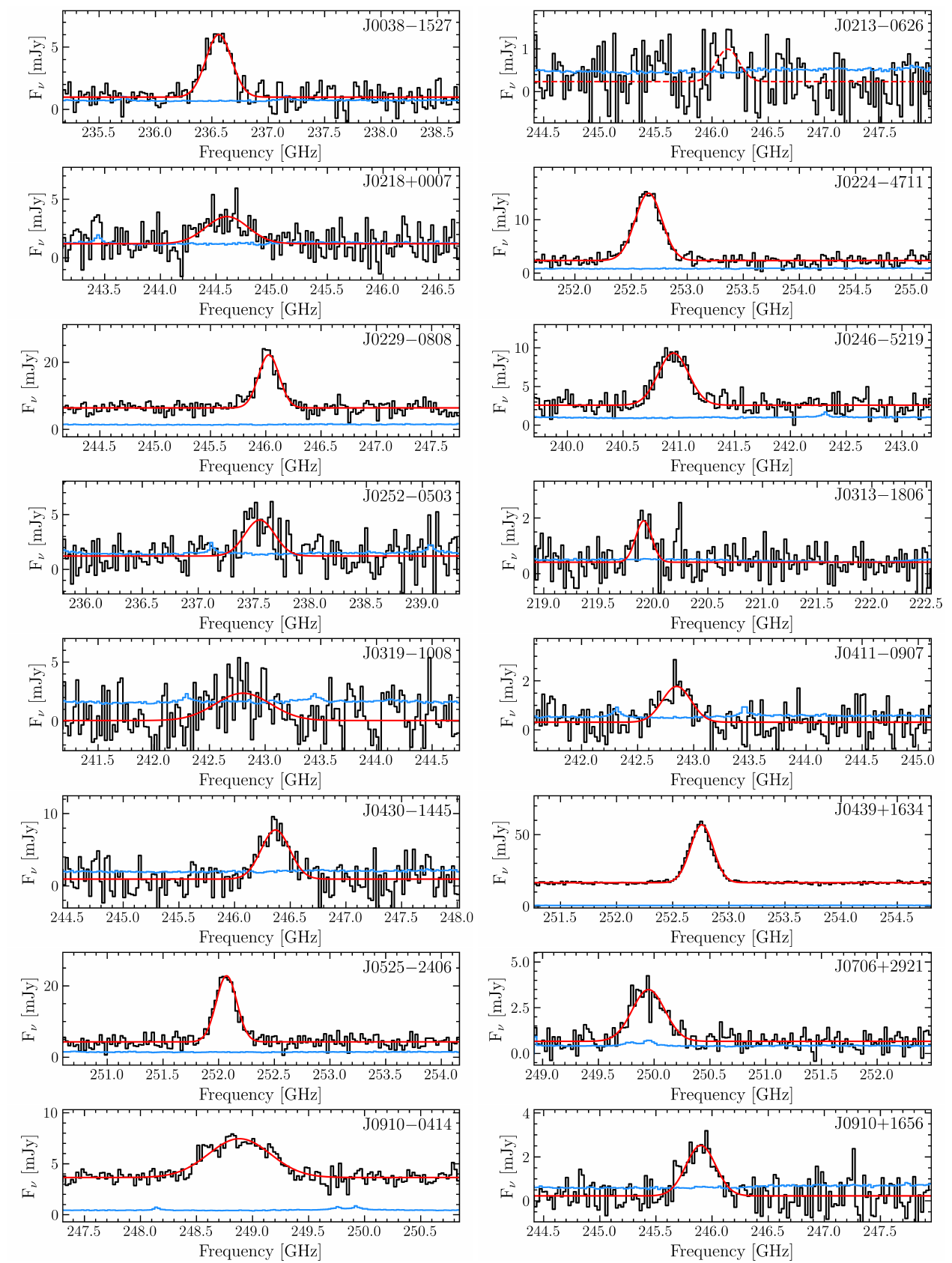}
\vspace{-5pt}
\caption{ALMA spectra of [\ion{C}{2}] and the underlying continuum of the quasars in our sample (black lines). The blue lines are $1\sigma$ error vectors. The best-fit Gaussian line+flat continuum models are shown as solid red lines. Dotted lines indicate the best fits in those cases where no significant line emission is detected.
\label{fig:spectra1}}
\end{figure*}

\addtocounter{figure}{-1}
\begin{figure*}
 \centering 
\includegraphics[width=0.98\textwidth]{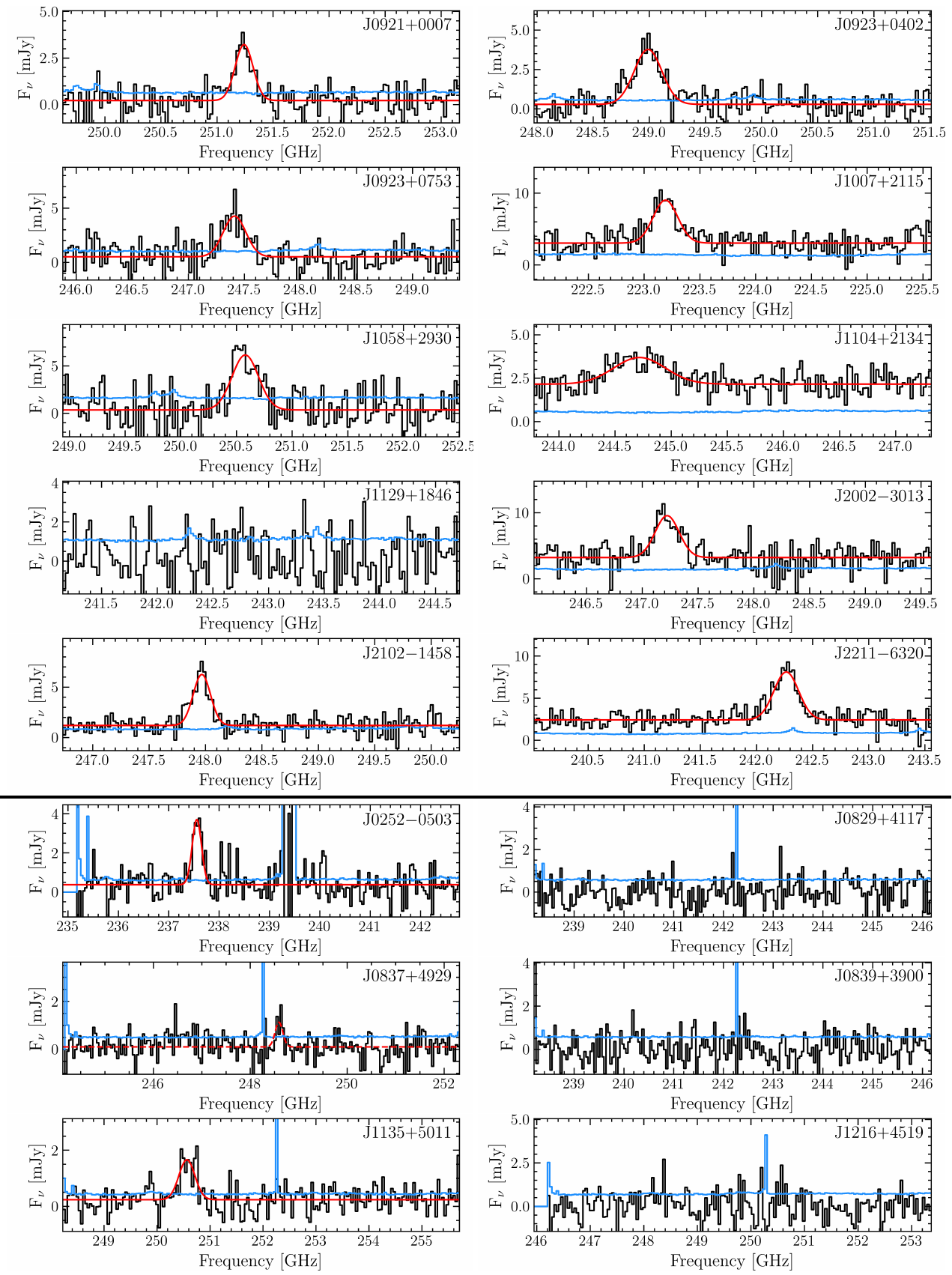}
\vspace{-5pt}
   \caption{Continued. The bottom six spectra are NOEMA observations while the rest are ALMA observations. }
   \label{fig:spectra2}
\end{figure*}

\subsection{NOEMA observation and data reduction}
Our NOEMA observations were obtained in the W18 cycle (program ID: W18EJ, PIs: B. Venemans and J. Yang). All quasars were observed using 10 antennas with the C array configuration during winter 2018. We used the wideband correlator {\tt PolyFiX}, which delivers two $\sim$7.7 GHz wide sidebands. The expected frequency of the [\ion{C}{2}] line of each quasar was tuned at the center of the inner baseband of the upper sideband. The total telescope time was designed to be $\sim4$ hours (on-source $\sim2.5$ hours) per target in order to reach $\sim$0.7 mJy/beam per 100 $\rm km ~s^{-1}$.  The detailed NOEMA observational information is also listed in Table \ref{tbl:basic}. 

The NOEMA data were analyzed using the {\tt mapping} software from the GILDAS\footnote{http://www.iram.fr/IRAMFR/GILDAS} suite. We use natural weighting and resampled the spectral axis in 50 $\rm km~s^{-1}$ wide channels. 
We extract the [\ion{C}{2}] 1D spectra from the peak positions for those [\ion{C}{2}] detected targets and extract the spectra from the optical positions for the [\ion{C}{2}] undetected targets. The extracted 1D spectra are shown in Figure \ref{fig:spectra1}.
Similar to the ALMA data reduction, the [\ion{C}{2}] line intensity map is collapsed from the data cube in the frequency range of $\pm1.4\sigma_{\rm line}$ from the line peak frequency (using values derived from \ion{Mg}{2} redshifts for [\ion{C}{2}] undetected sources).
The continuum maps are generated by collapsing all the line-free channels. 
The [\ion{C}{2}] line maps and the continuum contours of the NOEMA targets are also shown in Figure \ref {fig:map1}.

\begin{deluxetable*}{cccccccccccccccc}
\setlength{\tabcolsep}{2pt}
\tabletypesize{\scriptsize}
\tablecaption{The measured frequency, flux and width of the [\ion{C}{2}] line and the flux of underlying continuum.}\label{tbl:c2fit}
\tablehead{\colhead{Name} & \colhead{$\nu_{\rm obs}$} & \colhead{$z_{\rm [CII]}$} & \colhead{FWHM} & \colhead{$\sigma_{\rm [CII], Ho}$}  & \colhead{$F_{\rm [CII],1D}$} & \colhead{$F_{\nu,~ {\rm cont,1D}}$} & \colhead{$F_{\rm [CII],2D}$} & \colhead{$F_{\rm [CII],2D,corr}$} & \colhead{$F_{\nu,~ {\rm cont,2D}}$}
& \colhead{$v_{\rm proj}/2\sigma_{\rm int}$}       
} 
\startdata
                                             &    [GHz]                          &                       & [$\rm km~s^{-1}$]  & [$\rm km~s^{-1}$] & [$\rm Jy~km~s^{-1}$] &  [mJy]  & [$\rm Jy~km~s^{-1}$]  & [$\rm Jy~km~s^{-1}$]&  [mJy]  & \\
(1)     & (2)     & (3)     & (4)     & (5)     & (6)     & (7)     & (8)     & (9)     & (10)     & (11)     \\
\hline
J0038$-$1527 & 236.562$\pm$0.008 & 7.0340$\pm$0.0003 & 339$\pm$25 & 215$\pm$16 & 1.81$\pm$0.14 & 0.98$\pm$0.07 &  2.22$\pm$0.18 & 2.71$\pm$0.22 & 0.90$\pm$0.06 & 0.44$\pm$0.12 \\
J0213$-$0626 & 246.140$\pm$0.038 & 6.7214$\pm$0.0012 & 291$\pm$112 & 180$\pm$69 & 0.23$\pm$0.09 & 0.24$\pm$0.04 & 0.25$\pm$0.06\tablenotemark{a} & 0.30$\pm$0.07\tablenotemark{a} & 0.31$\pm$0.04 & -- \\
J0218$+$0007 & 244.599$\pm$0.040 & 6.7700$\pm$0.0013 & 548$\pm$120 & 379$\pm$83 & 1.35$\pm$0.30 & 1.21$\pm$0.13 & 1.29$\pm$0.28 & 1.53$\pm$0.33 & 1.05$\pm$0.06 & 0.28$\pm$0.09 \\
J0224$-$4711 & 252.656$\pm$0.004 & 6.5222$\pm$0.0001 & 334$\pm$15 & 211$\pm$9 & 4.49$\pm$0.20 & 2.36$\pm$0.09 & 4.24$\pm$0.12 & 5.11$\pm$0.14 & 2.12$\pm$0.06 & 0.41$\pm$0.10\\
J0229$-$0808 & 246.028$\pm$0.005 & 6.7249$\pm$0.0002 & 269$\pm$19 & 164$\pm$12 & 4.52$\pm$0.32 & 6.41$\pm$0.14 & 3.75$\pm$0.15 & 4.33$\pm$0.17 & 5.09$\pm$0.06 & 0.55$\pm$0.15\\
J0246$-$5219 & 240.951$\pm$0.009 & 6.8876$\pm$0.0003 & 400$\pm$29 & 261$\pm$19 & 2.87$\pm$0.21 & 2.59$\pm$0.09 & 2.26$\pm$0.13 & 2.72$\pm$0.16 & 1.41$\pm$0.04 & 0.38$\pm$0.13\\
J0252$-$0503 & 237.549$\pm$0.028 & 7.0006$\pm$0.0009 & 393$\pm$96 & 255$\pm$62 & 1.38$\pm$0.34 & 1.22$\pm$0.13 & 1.88$\pm$0.35 & 2.30$\pm$0.43 & 0.99$\pm$0.07 & 0.29$\pm$0.10\\
J0313$-$1806 & 219.914$\pm$0.016 & 7.6422$\pm$0.0006 & 235$\pm$56 & 141$\pm$34 & 0.37$\pm$0.09 & 0.41$\pm$0.04 & 0.38$\pm$0.08 & 0.45$\pm$0.09 & 0.48$\pm$0.04 & 0.28$\pm$0.13\\
J0319$-$1008 & 242.801$\pm$0.066 & 6.8275$\pm$0.0021 & 727$\pm$205 & 534$\pm$150 & 1.78$\pm$0.51 & 0.06$\pm$0.02 & 1.80$\pm$0.51 & 2.18$\pm$0.62 & 0.37$\pm$0.06 & -- \\
J0411$-$0907 & 242.848$\pm$0.022 & 6.8260$\pm$0.0007 & 371$\pm$67 & 238$\pm$43 & 0.58$\pm$0.11 & 0.31$\pm$0.05 & 1.52$\pm$0.34 & 1.83$\pm$0.41 & 0.22$\pm$0.06 & 0.39$\pm$0.13\\
J0430$-$1445 & 246.368$\pm$0.018 & 6.7142$\pm$0.0006 & 367$\pm$58 & 235$\pm$37 & 2.66$\pm$0.42 & 0.94$\pm$0.18 & 2.05$\pm$0.22 & 2.39$\pm$0.26 & 0.67$\pm$0.09 & 0.32$\pm$0.10\\
J0439$+$1634 & 252.759$\pm$0.001 & 6.5192$\pm$0.0001 & 283$\pm$4 & 174$\pm$2 & 12.15$\pm$0.16 & 16.52$\pm$0.08 & 13.16$\pm$0.17 & 15.77$\pm$0.20 & 16.62$\pm$0.11 & 0.45$\pm$0.18\\
J0525$-$2406 & 252.071$\pm$0.004 & 6.5397$\pm$0.0001 & 259$\pm$16 & 157$\pm$10 & 5.11$\pm$0.32 & 4.31$\pm$0.14 & 4.29$\pm$0.15 & 5.14$\pm$0.18 & 3.26$\pm$0.06 & 0.59$\pm$0.16\\
J0706$+$2921 & 249.950$\pm$0.012 & 6.6037$\pm$0.0003 & 413$\pm$36 & 271$\pm$24 & 1.24$\pm$0.11 & 0.66$\pm$0.06 & 1.70$\pm$0.22 & 2.06$\pm$0.27 & 0.78$\pm$0.05 & 0.46$\pm$0.17\\
J0910$+$1656 & 245.900$\pm$0.016 & 6.7289$\pm$0.0005 & 379$\pm$50 & 245$\pm$32 & 0.93$\pm$0.12 & 0.23$\pm$0.05 & 1.42$\pm$0.21 & 1.68$\pm$0.25 & 0.23$\pm$0.05 & 0.32$\pm$0.09\\
J0910$-$0414 & 248.883$\pm$0.011 & 6.6363$\pm$0.0003 & 783$\pm$40 & 584$\pm$30 & 3.17$\pm$0.16 & 3.66$\pm$0.06 & 3.21$\pm$0.17 & 3.83$\pm$0.20 & 3.38$\pm$0.04 & 0.33$\pm$0.14\\
J0921$+$0007 & 251.242$\pm$0.010 & 6.5646$\pm$0.0003 & 224$\pm$36 & 133$\pm$21 & 0.72$\pm$0.12 & 0.22$\pm$0.05 & 0.91$\pm$0.11 & 1.12$\pm$0.14 & 0.27$\pm$0.08 & 0.37$\pm$0.13\\
J0923$+$0402 & 248.988$\pm$0.009 & 6.6330$\pm$0.0003 & 350$\pm$36 & 223$\pm$23 & 1.30$\pm$0.13 & 0.29$\pm$0.05 & 1.58$\pm$0.18 & 1.87$\pm$0.21 & 0.47$\pm$0.11 & 0.29$\pm$0.10\\
J0923$+$0753 & 247.412$\pm$0.015 & 6.6817$\pm$0.0005 & 289$\pm$41 & 178$\pm$25 & 1.15$\pm$0.16 & 0.52$\pm$0.10 & 1.27$\pm$0.21 & 1.52$\pm$0.25 & 0.35$\pm$0.08 & 0.44$\pm$0.10\\
J1007$+$2115 & 223.191$\pm$0.014 & 7.5153$\pm$0.0005 & 349$\pm$56 & 222$\pm$36 & 2.22$\pm$0.36 & 3.04$\pm$0.12 & 1.78$\pm$0.15 & 2.14$\pm$0.18 & 1.91$\pm$0.07 & 0.35$\pm$0.11\\
J1058$+$2930 & 250.579$\pm$0.016 & 6.5846$\pm$0.0005 & 336$\pm$46 & 212$\pm$29 & 2.08$\pm$0.29 & 0.35$\pm$0.10 & 2.18$\pm$0.39 & 2.54$\pm$0.45 & 0.44$\pm$0.08 & 0.43$\pm$0.10\\
J1104$+$2134 & 244.718$\pm$0.029 & 6.7662$\pm$0.0009 & 664$\pm$95 & 478$\pm$68 & 1.08$\pm$0.16 & 2.16$\pm$0.06 & 1.35$\pm$0.21 & 1.62$\pm$0.25 & 1.99$\pm$0.04 & 0.20$\pm$0.07\\
J1129$+$1846 & -- & -- & -- & -- & -- & -- & $<0.12$\tablenotemark{b} & $<0.14$\tablenotemark{b} & 0.18$\pm$0.08\tablenotemark{c}  & --\\ 
J2002$-$3013 & 247.223$\pm$0.013 & 6.6876$\pm$0.0004 & 308$\pm$36 & 192$\pm$22 & 2.07$\pm$0.25 & 3.22$\pm$0.15 & 1.68$\pm$0.17 & 1.99$\pm$0.20 & 1.88$\pm$0.05 & 0.42$\pm$0.11\\
J2102$-$1458 & 247.967$\pm$0.008 & 6.6645$\pm$0.0002 & 219$\pm$29 & 130$\pm$17 & 1.18$\pm$0.16 & 1.20$\pm$0.08 & 1.20$\pm$0.11 & 1.38$\pm$0.13 & 0.95$\pm$0.06 & 0.37$\pm$0.11\\
J2211$-$6320 & 242.265$\pm$0.010 & 6.8449$\pm$0.0003 & 321$\pm$31 & 201$\pm$19 & 1.94$\pm$0.19 & 2.42$\pm$0.08 & 1.66$\pm$0.12 & 1.99$\pm$0.14 & 1.67$\pm$0.04 & 0.39$\pm$0.08\\
\hline
J0252$-$0503 & 237.557$\pm$0.012 & 7.0003$\pm$0.0004 & 275$\pm$35 & 168$\pm$21 & 0.97$\pm$0.12 & 0.37$\pm$0.05 & 1.35$\pm$0.11\tablenotemark{d} & 1.63$\pm$0.13\tablenotemark{d} & 0.66$\pm$0.04 & --\\ 
J0829$+$4117 & -- & -- & -- & -- & -- & -- & $<0.33$\tablenotemark{b} & $<0.40$\tablenotemark{b}  & $<0.14$\tablenotemark{b}  & -- \\ 
J0837$+$4929 & 248.591$\pm$0.032 & 6.6452$\pm$0.0010 & 225$\pm$109 & 134$\pm$65 & 0.25$\pm$0.12 & 0.10$\pm$0.03 & 0.58$\pm$0.08\tablenotemark{d} & 0.70$\pm$0.10\tablenotemark{d} & 0.18$\pm$0.03 & --\\ 
J0839$+$3900 & -- & -- & -- & -- & -- & -- & $<0.30$\tablenotemark{b} & $<0.36$\tablenotemark{b} & $<0.15$\tablenotemark{b} & --\\ 
J1135$+$5011 & 250.563$\pm$0.027 & 6.5851$\pm$0.0008 & 425$\pm$75 & 280$\pm$49 & 0.64$\pm$0.11 & 0.24$\pm$0.04 & 1.46$\pm$0.11\tablenotemark{d} & 1.76$\pm$0.13\tablenotemark{d} & 0.33$\pm$0.03 & --\\
J1216$+$4519 & -- & -- & -- & -- & -- & -- & $<0.30$\tablenotemark{b} & $<0.36$\tablenotemark{b} & $<0.12$\tablenotemark{b}  & --
\enddata
{{\bf Notes:} (1) quasar name, (2) observed [\ion{C}{2}] central frequency, (3) redshift from [\ion{C}{2}] line, (4) FWHM of the [\ion{C}{2}] lines derived from 1D Gaussian fitting, (5) [\ion{C}{2}] line $\sigma$ derived following \cite{Ho07} by correcting the turbulent broadening and the inclination factor, (6) [\ion{C}{2}] line flux measured from 1D Gaussian fitting, (7) continuum flux measured by fitting the 1D spectra, (8) [\ion{C}{2}] line flux measured from the 2D maps, (9) [\ion{C}{2}] line flux measured from the 2D maps after correcting the missing flux from the wings of the line that are not included in the 2D maps, (10) continuum flux measured from the 2D continuum maps, (11) ratio of half the projected velocity gradient (without correcting inclination) to the source-integrated velocity dispersion (see \S \ref{subsec:morphology}).}
\tablenotetext{a}{Flux was estimated from a circle centered at the quasar position with a diameter of 1\farcs5.}
\tablenotetext{b}{3$\sigma$ upper limit.}
\tablenotetext{c}{Flux was estimated from the taper image with a beam size of 1\farcs0.}
\tablenotetext{d}{Flux was estimated from a circle centered at the quasar position with a diameter of 3\farcs0.}

\end{deluxetable*}

\section{Measurements and Correlations} \label{sec:rlt}

\begin{deluxetable*}{cccccccccccc}
\setlength{\tabcolsep}{2pt}
\tabletypesize{\scriptsize}
\tablecaption{The luminosity, SFR and mass measurements}\label{tbl:measure}
\tablehead{\colhead{Name} & \colhead{$L_{\rm FIR, graybody}$} & \colhead{$L_{\rm TIR, graybody}$} & \colhead{$L_{\rm TIR, Haro11}$} & \colhead{$L_{\rm [CII]}$} & 
                 \colhead{$\rm SFR_{graybody}$} & \colhead{$\rm SFR_{Haro11}$} & \colhead{$\rm SFR_{[CII]}$} & \colhead{$M_{\rm dust}$} & \colhead{$M_{\rm C^+}$} & \colhead{$M_{\rm dyn}$\tablenotemark{a}} & \colhead{$M_{\rm dyn, N21}$\tablenotemark{b}} }
\startdata
        &   [$10^{12}~L_\odot$]  &   [$10^{12}~L_\odot$]  &   [$10^{12}~L_\odot$] &   [$10^{9}~L_\odot$] &   [$\rm M_\odot~yr^{-1}$] &  [$\rm M_\odot~yr^{-1}$] &  [$\rm M_\odot~yr^{-1}$]  &  [$10^8~M_\odot$] & [$10^6~M_\odot$] & [$10^{11}~M_\odot$] & [$10^{11}~M_\odot$]\\
(1)     & (2)     & (3)     & (4)     & (5)     & (6)     & (7)     & (8)     & (9)     & (10)       & (11)       & (12)     \\
\hline
J0038$-$1527 & 2.49$\pm$0.17 & 3.52$\pm$0.23 & 5.13$\pm$0.34 & 3.21$\pm$0.26 & 522$\pm$34 & 981$\pm$65 & 199-1246 &  1.60$\pm$0.11 &  9.58$\pm$0.78 & $2.00_{-0.77}^{+2.15}$ & $0.89_{-0.34}^{+0.95}$\\
J0213$-$0626 & 0.79$\pm$0.10 & 1.12$\pm$0.14 & 1.65$\pm$0.21 & 0.33$\pm$0.08 & 165$\pm$21 & 316$\pm$40 &    13-86 &  0.51$\pm$0.07 &  0.99$\pm$0.23 & -- & --\\
J0218$+$0007 & 2.71$\pm$0.15 & 3.82$\pm$0.22 & 5.65$\pm$0.32 & 1.71$\pm$0.37 & 567$\pm$32 &1081$\pm$61 &   95-594 &  1.74$\pm$0.10 &  5.12$\pm$1.11 & -- & --\\
J0224$-$4711 & 5.09$\pm$0.14 & 7.17$\pm$0.20 &10.75$\pm$0.30 & 5.43$\pm$0.15 &1065$\pm$30 &2057$\pm$58 & 370-2316 &  3.26$\pm$0.09 & 16.20$\pm$0.46 & $2.09_{-0.92}^{+3.01}$ & $0.93_{-0.41}^{+1.34}$\\
J0229$-$0808 &13.04$\pm$0.15 &18.39$\pm$0.22 &27.24$\pm$0.32 & 4.81$\pm$0.19 &2730$\pm$32 &5212$\pm$61 & 321-2006 &  8.35$\pm$0.10 & 14.34$\pm$0.57 & $1.00_{-0.44}^{+1.46}$ & $0.45_{-0.20}^{+0.65}$\\
J0246$-$5219 & 3.75$\pm$0.11 & 5.29$\pm$0.15 & 7.77$\pm$0.22 & 3.13$\pm$0.18 & 785$\pm$22 &1487$\pm$42 & 193-1210 &  2.40$\pm$0.07 &  9.34$\pm$0.54 & $1.65_{-0.63}^{+1.76}$ & $0.73_{-0.28}^{+0.78}$\\
J0252$-$0503 & 2.73$\pm$0.19 & 3.85$\pm$0.27 & 5.62$\pm$0.40 & 2.71$\pm$0.50 & 571$\pm$40 &1075$\pm$76 & 163-1020 &  1.75$\pm$0.12 &  8.09$\pm$1.51 & -- & --\\
J0313$-$1806 & 1.30$\pm$0.11 & 1.83$\pm$0.15 & 2.66$\pm$0.22 & 0.60$\pm$0.13 & 272$\pm$22 & 508$\pm$42 &   27-171 &  0.78$\pm$0.09 &  2.38$\pm$0.65 & -- & --\\
J0319$-$1008 & 0.97$\pm$0.16 & 1.37$\pm$0.22 & 2.02$\pm$0.33 & 2.47$\pm$0.70 & 203$\pm$33 & 386$\pm$62 &  146-916 &  0.62$\pm$0.10 &  7.38$\pm$2.09 & -- & --\\
J0411$-$0907 & 0.58$\pm$0.16 & 0.81$\pm$0.22 & 1.20$\pm$0.33 & 2.08$\pm$0.46 & 120$\pm$32 & 229$\pm$62 &  119-745 &  0.37$\pm$0.10 &  6.20$\pm$1.39 & -- & --\\
J0430$-$1445 & 1.71$\pm$0.23 & 2.41$\pm$0.32 & 3.58$\pm$0.48 & 2.65$\pm$0.28 & 358$\pm$48 & 684$\pm$91 &  159-994 &  1.10$\pm$0.15 &  7.91$\pm$0.85 & -- & --\\
J0525$-$2406 & 7.89$\pm$0.15 &11.13$\pm$0.20 &16.66$\pm$0.31 & 5.48$\pm$0.19 &1653$\pm$30 &3188$\pm$58 & 374-2343 &  5.06$\pm$0.09 & 16.35$\pm$0.57 & $0.33_{-0.02}^{+0.02}$ & $0.14_{-0.01}^{+0.01}$ \\
J0706$+$2921 & 1.91$\pm$0.12 & 2.70$\pm$0.17 & 4.03$\pm$0.26 & 2.23$\pm$0.29 & 400$\pm$25 & 770$\pm$49 &  129-808 &  1.23$\pm$0.08 &  6.64$\pm$0.86 & $2.17_{-0.78}^{+2.46}$ & $0.96_{-0.34}^{+1.10}$\\
J0829$+$4117 &       $<$0.33 &       $<$0.47 &       $<$0.71 &       $<$0.45 &      $<$70 &     $<$135 &$<$19-122 &$<$0.21&$<$1.34&   -- & --\\
J0837$+$4929 & 0.42$\pm$0.07 & 0.59$\pm$0.10 & 0.88$\pm$0.15 & 0.76$\pm$0.11 & 87$\pm$14  & 168$\pm$28 &   36-229 &  0.27$\pm$0.04 &  2.28$\pm$0.33 & -- & -- \\
J0839$+$3900 &       $<$0.37 &       $<$0.52 &       $<$0.78 &       $<$0.42 &      $<$77 &     $<$148 &$<$17-111 &$<$0.24&$<$1.24&   -- & --\\
J0910$+$1656 & 0.59$\pm$0.13 & 0.83$\pm$0.18 & 1.23$\pm$0.27 & 1.87$\pm$0.28 & 123$\pm$26 & 234$\pm$51 &  105-658 &  0.38$\pm$0.08 &  5.58$\pm$0.82 & -- & --\\
J0910$-$0414 & 8.43$\pm$0.10 &11.89$\pm$0.14 &17.70$\pm$0.21 & 4.17$\pm$0.22 &1766$\pm$20 &3387$\pm$40 & 271-1697 &  5.40$\pm$0.06 & 12.44$\pm$0.66 & $16.78_{-10.26}^{+121.38}$ & $7.46_{-4.56}^{+53.95}$\\
J0921$+$0007 & 0.66$\pm$0.20 & 0.93$\pm$0.28 & 1.39$\pm$0.41 & 1.20$\pm$0.14 & 138$\pm$40 & 266$\pm$78 &   62-389 &  0.42$\pm$0.13 &  3.58$\pm$0.43 & -- & --\\
J0923$+$0402 & 1.16$\pm$0.27 & 1.64$\pm$0.38 & 2.44$\pm$0.57 & 2.04$\pm$0.23 & 243$\pm$57 &467$\pm$109 &  116-729 &  0.74$\pm$0.17 &  6.08$\pm$0.69 & -- & --\\
J0923$+$0753 & 0.88$\pm$0.20 & 1.24$\pm$0.28 & 1.85$\pm$0.42 & 1.67$\pm$0.28 & 184$\pm$42 & 353$\pm$80 &   92-577 &  0.56$\pm$0.13 &  4.99$\pm$0.83 & -- & --\\
J1007$+$2115 & 5.05$\pm$0.19 & 7.12$\pm$0.26 &10.39$\pm$0.38 & 2.79$\pm$0.23 &1057$\pm$38 &1988$\pm$72 & 168-1055 &  3.23$\pm$0.12 &  8.32$\pm$0.70 & -- & --\\
J1058$+$2930 & 1.08$\pm$0.20 & 1.52$\pm$0.28 & 2.27$\pm$0.41 & 2.74$\pm$0.49 & 226$\pm$41 & 435$\pm$79 & 165-1031 &  0.69$\pm$0.13 &  8.16$\pm$1.46 & $1.83_{-0.82}^{+3.57}$ & $0.81_{-0.36}^{+1.58}$\\
J1104$+$2134 & 5.10$\pm$0.10 & 7.20$\pm$0.14 &10.65$\pm$0.21 & 1.82$\pm$0.28 &1069$\pm$21 &2039$\pm$40 &  102-638 &  3.27$\pm$0.07 &  5.43$\pm$0.85 & -- & --\\
J1129$+$1846 & 0.44$\pm$0.19 & 0.61$\pm$0.27 & 0.92$\pm$0.41 &       $<$0.16 &  91$\pm$40 & 175$\pm$78 &  $<$5-35 &  0.30$\pm$0.13 &$<$0.50&   -- & --\\
J1135$+$5011 & 0.75$\pm$0.07 & 1.06$\pm$0.10 & 1.60$\pm$0.15 & 1.90$\pm$0.14 & 157$\pm$14 & 305$\pm$27 &  107-669 &  0.48$\pm$0.04 &  5.66$\pm$0.42 & -- & --\\
J1216$+$4519 &       $<$0.28 &       $<$0.39 &       $<$0.59 &        $<$0.39&      $<$58 &     $<$112 &$<$16-104 &$<$0.18&$<$1.17&   -- & --\\
J2002$-$3013 & 4.73$\pm$0.13 & 6.67$\pm$0.18 & 9.91$\pm$0.26 & 2.20$\pm$0.22 & 990$\pm$26 &1897$\pm$50 &  127-796 &  3.03$\pm$0.08 &  6.56$\pm$0.66 & $0.57_{-0.22}^{+0.69}$ & $0.25_{-0.10}^{+0.31}$\\
J2102$-$1458 & 2.38$\pm$0.15 & 3.35$\pm$0.21 & 4.98$\pm$0.31 & 1.51$\pm$0.14 & 497$\pm$31 & 953$\pm$60 &   81-511 &  1.52$\pm$0.10 &  4.50$\pm$0.41 & -- & --\\
J2211$-$6320 & 4.44$\pm$0.11 & 6.26$\pm$0.15 & 9.21$\pm$0.22 & 2.27$\pm$0.16 & 929$\pm$22 &1761$\pm$42 &  132-828 &  2.84$\pm$0.07 &  6.77$\pm$0.49 & -- & --
\enddata
{{\bf Notes:} (1) quasar name, (2) far-infrared luminosity estimated from graybody fitting, (2) total infrared luminosity estimated from graybody fitting, (4) total infrared luminosity estimated from Haro11 template, (5) [\ion{C}{2}] line luminosity, (6) star formation rate estimated from $L_{\rm TIR, graybody}$, (7) star formation rate estimated from $L_{\rm TIR, Haro11}$, (8) star formation rate estimated from [\ion{C}{2}], (9) dust mass, (10) the mass of singly ionized carbon.}
\tablenotetext{a}{Dynamical mass measured from Equation \ref{eq:dynamical}.}
\tablenotetext{b}{Dynamical mass measured from Equation \ref{eq:dynamical_N21}.}
\end{deluxetable*}

\begin{figure}
\centering
\includegraphics[width=0.48\textwidth]{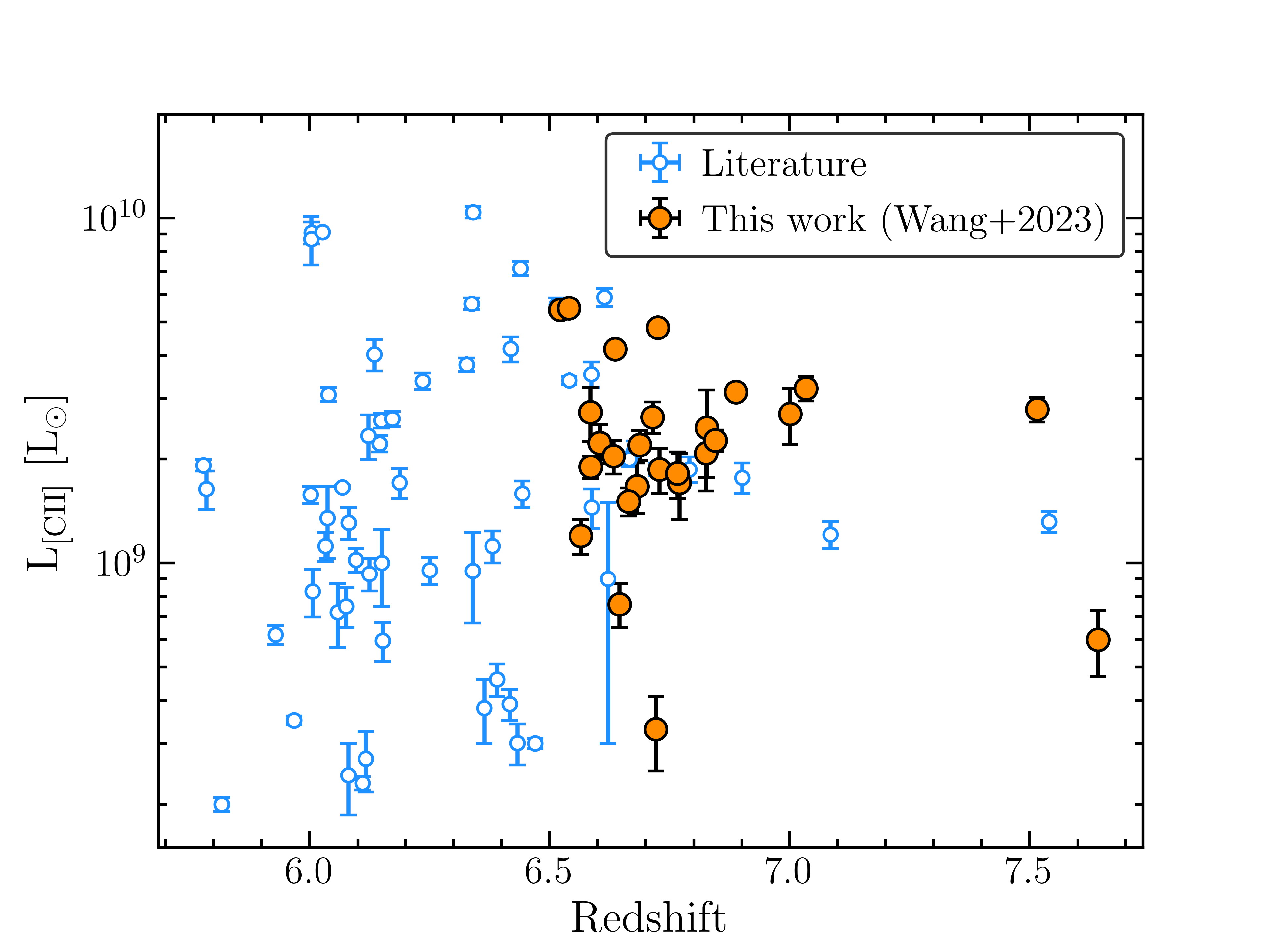}
\caption{The [\ion{C}{2}] luminosity as a function of redshift for the quasars from our survey (orange points) and all other quasars with [\ion{C}{2}] detections at $z>5.7$ (blue circles). 
Our ALMA and NOEMA surveys detect the [\ion{C}{2}] emission of 26 $z>6.5$ quasars, representing 67\% of all quasars at $z>6.5$ with [\ion{C}{2}] detections. Note that the
brightest object is the lensed quasar J0439+1634 without correcting the lensing magnification. 
\label{fig:Lc2}}
\end{figure}

\subsection{[\ion{C}{2}] and Infrared Luminosity Measurements}
As described in \S \ref{sec:obs}, the extracted spectra are shown in Figure \ref{fig:spectra1}. We fit the spectra by assuming a flat dust continuum emission and a Gaussian profile for the [\ion{C}{2}] emission line. 
The derived line peak frequency, full width at half maximum (FWHM), and the line flux as well as their associated $1\sigma$ uncertainties are listed in Table \ref{tbl:c2fit}. 
We consider a line detection to be significant if its integrated line flux is $>3\sigma$ and marginal if its integrated line flux is between $2\sigma$ and $3\sigma$.  
Out of 31 targeted objects, only four quasars 
(J0829+4117, J0839+3900, J1129+1846, and J1216+4519) are undetected in [\ion{C}{2}] and two quasars have marginal line detections at a 2.6$\sigma$ level (J0213$-$0626) and a 2.1$\sigma$ level (J0837+4929), respectively. 
Three (J0829+4117, J0839+3900, and J1216+4519) out of the four [\ion{C}{2}]-undetected quasars are from the NOEMA observations, which have a depth that is two times shallower than that of the ALMA observations. 
The FIR continuum flux and flux uncertainty measured from the 1D spectral fitting are also listed in Table \ref{tbl:c2fit}. All objects observed with ALMA, except for J1129+1846, are detected in the dust continuum at high significance. 
Similar to the  [\ion{C}{2}] results, the FIR continua of  J0829+4117, J0839+3900, and J1216+4519 are not detected by NOEMA. 
J1129+1846, the only known radio-loud quasar in our sample, is intrinsically faint in both [\ion{C}{2}] and FIR \citep[see also][]{Khusanova22}. 

We also derived the line flux and the size of the [\ion{C}{2}] emitting region for the ALMA-detected objects by fitting the [\ion{C}{2}] line maps shown in Figure \ref{fig:map1}. The fit is performed with the {\tt imfit} task within CASA by assuming a 2D Gaussian profile with the centroid, integrated flux, major and minor axes, and the position angle as free parameters. We successfully fit all the detected [\ion{C}{2}] emission except for J0213$-$0626 because of the low significance of its detection. Since the [\ion{C}{2}] maps are created by integrating over $\pm1.4\sigma_{\rm line}$, which encloses $\sim83$\% of the total line flux for a Gaussian profile, we take this factor into account when computing the total line flux \citep[see also][]{Decarli18}. The line flux and the uncertainty measured from the 2D fitting are also listed in Table \ref{tbl:c2fit}. For J0213--0626, we use the line flux calculated from the intensity map with a 1\farcs5 diameter aperture after correcting for the missing flux. 
We note that the line fluxes measured from the 2D map could be different from those measured in 1D spectra in some cases, especially when the galaxy is very extended (e.g., J0411-0907). This is mainly because the 1D spectral extraction could not enclose all flux from the quasar host galaxies. Therefore, we use the 2D line fluxes in the following analyses \citep[see also][]{Decarli18}.
For the NOEMA-detected objects, we calculate the [\ion{C}{2}] flux from the intensity map with a 3\farcs0 diameter aperture after correcting for the missing flux. 
The 3$\sigma$ upper limits are reported in Table \ref{tbl:c2fit} for the [\ion{C}{2}]-undetected targets. 
We then measure the line luminosity following
\begin{equation}\label{eq_line_lum}
\frac{L_{\rm [CII]}}{\rm L_\odot}=1.04\times10^{-3}\, \frac{F_{\rm [CII]}}{\rm Jy\,km\,s^{-1}}\,\frac{\nu_{\rm obs}}{\rm GHz}\, \left(\frac{D_{\rm L}}{\rm Mpc}\right)^2,
\end{equation}
where $F_{\rm [CII]}$ is the total line flux, $\nu_{\rm obs}$ is the observed frequency of the [\ion{C}{2}] line, and $D_{\rm L}$ is the luminosity distance. 
The [\ion{C}{2}] line luminosities, $L_{\rm [CII]}$, of all detected quasar host galaxies in our sample are listed in Table \ref{tbl:measure} and they span a range of $L_{\rm [CII]}=(0.3-5.5)\times10^9~L_\odot$ with the majority of them having $L_{\rm [CII]}>10^9~L_\odot$. 

\begin{figure*}
\centering
\includegraphics[width=0.98\textwidth]{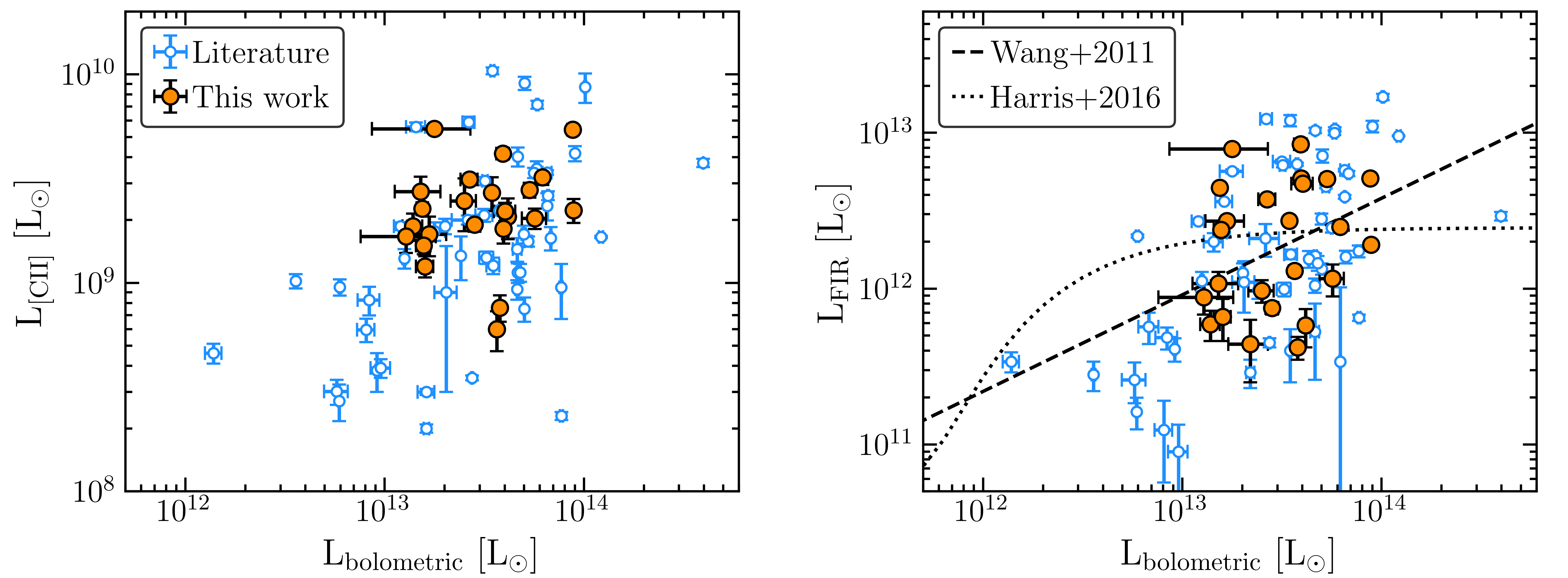}
\caption{The relations between quasar bolometric luminosity and the [\ion{C}{2}] luminosity ($left$) as well as the far-infrared luminosity ($right$) at $z>5.7$, respectively. The Spearman test gives a correlation coefficient of $\rho=0.43$ and a chance probability of $p=1.7\times10^{-4}$ for the $L_{\rm  [CII]}- L_{\rm bolometric}$ relation, and $\rho=0.48$,  $p=7.4\times10^{-6}$ for the $L_{\rm FIR}- L_{\rm bolometric}$ relation, respectively. A moderate correlation exits for both relations, though both relations have  significant scatter.  
\label{fig:lbol}}
\end{figure*}

To compare with observations available in the literature, we collected the properties of both [\ion{C}{2}] and FIR continua of $z>5.7$ quasars. There are 65 $z>5.7$ quasars with [\ion{C}{2}] and FIR continuum observations in the literature at the time of doing our analyses. Among them, 58 systems are detected in the [\ion{C}{2}] line, with 13 at $z>6.5$. Since most of the non-detections are caused by shallower observations, we only include the [\ion{C}{2}]-detected sources in the following analyses. 
Figure \ref{fig:Lc2} shows the $L_{\rm [CII]}$ distribution of quasar hosts detected from this survey and all other known $z>5.7$ quasars with published $L_{\rm [CII]}$ measurements. Our [\ion{C}{2}] survey significantly increases the number of [\ion{C}{2}] detections in quasar host galaxies at $z>5.7$ and more than doubles the number of [\ion{C}{2}] detections in quasar host galaxies at $z>6.5$. 
This program provides a complete coverage in [\ion{C}{2}] of the most luminous quasars (i.e., $M_{1450}<-25.0$) known at $z>6.5$. 
In the left panel of Figure \ref{fig:lbol}, we show the scatter plot between $L_{\rm [CII]}$ and $L_{\rm bolometric}$ for both our sample and the literature sample. In order to examine whether these two qualities correlate with each other, we did a Spearman test on measurements from both this work and literature. The Spearman test gives a correlation coefficient of $\rho=0.43$ and a chance probability of $p=1.7\times10^{-4}$, indicating a moderate correlation between these two quantities though a large scatter is present. This is consistent with what has been found with a smaller quasar sample at $z\sim6$ \citep[e.g.][]{Decarli18,Izumi19}. This suggests that more luminous quasars in general hosted by galaxies with brighter [\ion{C}{2}] emission, however, the correlation could be biased since the [\ion{C}{2}] observation were only performed for UV luminous quasars (i.e., UV faint quasars with bright [\ion{C}{2}] emission is missed by current surveys). 

Since most of the ALMA-observed targets are marginally resolved (i.e., emission extent is $\gtrsim~ 1.5\times$ the beam size), we fit their continuum maps with {\tt imfit} to derive the continuum flux and size. 
The measured dust continuum fluxes are listed in Table \ref{tbl:c2fit}. J1129+1846 is only marginally detected in the 2D continuum map and we estimate the continuum flux from a taper image (with a 1\farcs0 beam) without measuring the size. 
The continua of the quasar host galaxies observed by NOEMA are not spatially resolved, therefore we use the peak values from the 2D continuum maps as their continuum fluxes. 
The FIR rest-frame 42.5--122.5 $\mu$m luminosities, $L_{\rm FIR}$, and the total infrared (TIR, rest-frame 8--1000 $\mu$m) luminosities, $L_{\rm TIR}$, are then estimated from the 2D continuum flux measurements. 
In order to infer the FIR and TIR luminosities, we used two different models: a galaxy template from Haro 11 which has a relatively low metallicity and is thought to be a good analogue of high-redshift galaxies \citep[e.g.][]{Lyu16}, and a modified blackbody in the optically thin regime, i.e., assuming the dust optical depth $\tau_{\rm dust}\ll1$  \citep[e.g.][]{Beelen06,Venemans18}. For the case of the Haro 11 template, we simply scale the template to the measured continuum flux of each target and then derive $L_{\rm FIR}$ and $L_{\rm TIR}$ by integrating the galaxy template from 42.5 to 122.5 $\mu$m and from 8 to 1000 $\mu$m, respectively. For the case of the graybody model, we assume a dust temperature of $T_{\rm dust}=47$ K and an emissivity index of $\beta=1.6$ \citep{Beelen06}. Since the cosmic microwave background reduces the flux densities that we can measure from sources at high redshift, we also take the CMB effect into account, following \cite{daCunha13} when fitting the graybody. The measured infrared luminosities for both cases are listed in Table \ref{tbl:measure}. 
The $L_{\rm TIR}$ measured using the galaxy template from Haro 11 is in general 1.5 times higher than that was measured from the graybody fitting. Since the modified graybody has been more broadly used in the literature, we only use the luminosities derived from the graybody fitting in the following analyses. There is no precise magnification measurement in FIR of the gravitationally lensed quasar, J0439+1634, thus we only report the flux measurements of this source and discard the luminosity and size measurements in the following analyses. We refer the readers to \cite{Yang19b} and \cite{Yue21} for a detailed characterization of the host galaxy of this quasar. 

The $L_{\rm FIR}$ (measured from the graybody fitting) of our targets has a broad range between $\sim5\times10^{11}~L_\odot$ and $\sim1.3\times10^{13}~L_\odot$, similar to what has been found for slightly lower redshift quasar host galaxies \citep[e.g.][]{Venemans20} but significantly brighter than $z\gtrsim6$ optically-selected star-forming galaxies \citep[e.g.][]{Smit18, Bouwens21}. 
In the right panel of Figure \ref{fig:lbol}, we show the scatter plot between $L_{\rm FIR}$ and $L_{\rm bolometric}$. The Spearman test gives a correlation coefficient of $\rho=0.48$ and a chance probability of $p=7.4\times10^{-6}$. This indicates a mild positive correlation exists between $L_{\rm FIR}$ and $L_{\rm bolometric}$ although with significant scatter. 
This is also consistent with similar studies at slightly lower redshifts by \cite{Izumi18}. We note that \cite{Venemans20} claimed a weaker or even no relationships between $L_{\rm FIR}$ and $L_{\rm bolometric}$, but that could be caused by the lack of low-luminosity objects in their sample.
The $L_{\rm bolometric}$ is a proxy for ongoing black hole growth while the $L_{\rm FIR}$ and $L_{\rm [CII]}$ are good tracers of star formation in the quasar host galaxies. The mild positive $L_{\rm FIR}-L_{\rm bolometric}$ and $L_{\rm [CII]}-L_{\rm bolometric}$ correlations of these quasars indicate that the more active SMBHs in the early Universe generally resides in galaxies with more active star forming activities. However, the large scatters in both relations could be caused by a variety of effects, including different timescales for black hole accretion and star formation. However, we noted that these correlations could be biased since current observations were targeted for UV luminous quasars and a population of UV faint quasars with strong activate star formation activities could be missed by our survey.

\subsection{[\ion{C}{2}] and FIR Continuum Size Measurements}\label{subsec:size}
\begin{figure}
\centering
\includegraphics[width=0.48\textwidth]{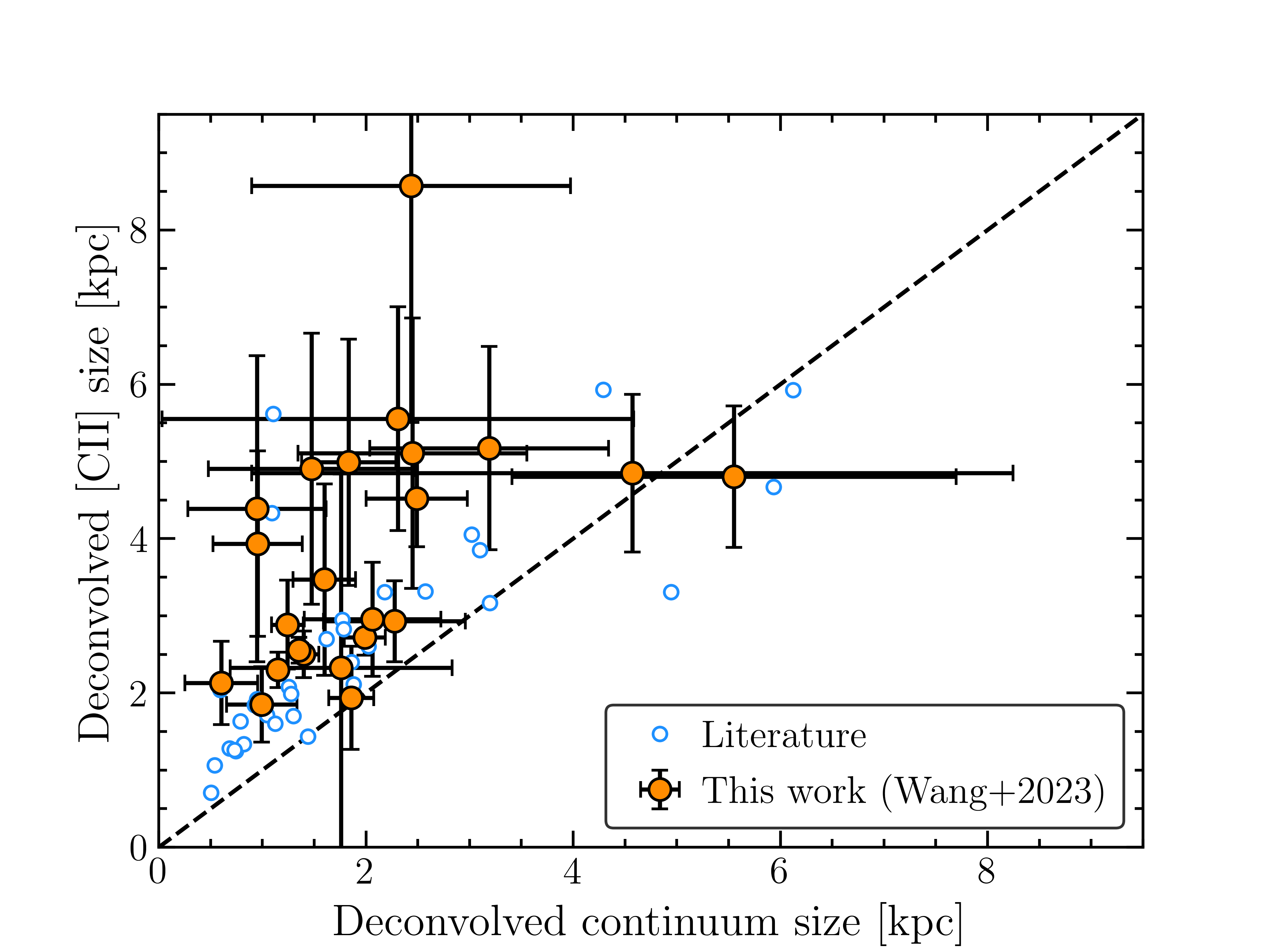}
\caption{Comparison of the sizes of the continuum and [\ion{C}{2}] emissions. The sizes are measured after deconvolution with the beam. The black dashed line denotes the one-to-one relation. In general, the continuum emission is more compact than the [\ion{C}{2}] emission.
\label{fig:size}}
\end{figure}

As described above, we fit the 2D maps of both [\ion{C}{2}] and continuum emission with {\tt imfit} for the sources detected with ALMA. The derived sizes are listed in Table \ref{tbl:size}. 
The ratio between the [\ion{C}{2}] major axis size and the beam major axis size ranges from $\sim$1.3 to $\sim$3.0 (see the column S/$\rm B_{[CII]}$ of Table \ref{tbl:size}), suggesting that the [\ion{C}{2}] emission regions of all quasars detected by ALMA are marginally resolved. The deconvolved sizes of the [\ion{C}{2}] range from 0\farcs44$\times$0\farcs34 (or 2.20 kpc $\times$ 1.70 kpc) to 1\farcs85$\times$1\farcs41 (or 9.82 kpc $\times$ 7.48 kpc) with a median major axis FWHM of 0\farcs7 or 3.5 kpc. The ratio between the continuum emitting region size and the beam size ranges from $\sim$1.1 to $\sim$2.4 (see the column S/$\rm B_{cont}$ of Table \ref{tbl:size}). 
The deconvolved major axis size of the continuum emitting region ranging from 0\farcs20$\times$0\farcs16 (or 1.06 kpc $\times$ 0.85 kpc) to 1\farcs34$\times$0\farcs79 (or 7.23 kpc $\times$ 4.26 kpc) with a median major axis FWHM of 0\farcs4 or 2.1 kpc. 
In Figure \ref{fig:size}, we show the comparison of the [\ion{C}{2}] effective sizes and continuum effective sizes, where the effective sizes were derived with $\rm \sqrt{FWHM_{\rm maj}FWHM_{\rm min}}$.
In this plot, we also include all effective size measurements collected from the literature. In general, the continuum emission is more concentrated than the [\ion{C}{2}] emission, consistent with previous studies of both quasar host galaxies \citep[e.g.][]{Venemans20, Novak20, Walter22} and high-redshift star-forming galaxies \citep[e.g.][]{Capak15, Fujimoto20}.
This could be caused by the fact that the radiation in the central region closer to the quasar is more intense which yields a lower $L_{\rm [CII]}$/$L_{\rm FIR}$ (i.e., the so-called [\ion{C}{2}] deficit). In \S \ref{subsec:deficit}, we will discuss the [\ion{C}{2}] deficit in more detail.

Studies of the dynamical mass of quasar host galaxies with low-resolution [\ion{C}{2}] observations usually assume an inclination angle \citep[e.g. $i\sim55^\circ$;][]{Willott15,Decarli18} and a disk size for the hosts \citep[e.g. $D\sim4.5$ kpc;][]{Wang16}. These values are derived from the early ALMA observations of five quasar host galaxies at $z\sim6$ \citep{Wang13}. Using the spatially resolved observations from ALMA, we can update the median size and inclination angle of quasar host galaxies based on a much larger quasar sample (23 quasars from this work and 49 quasars from the literature). Following previous work \citep[e.g.,][]{Wang13, Willott15}, we estimate the galaxy diameter $D$ to be 1.5$\times$ the FWHM major axis from the [\ion{C}{2}] intensity map (i.e., full width at 20\% of the peak intensity for a Gaussian profile), and we estimate the inclination angle $i$ from the ratio of minor ($a_{\rm min}$) and major ($a_{\rm maj}$) axes according to $i = {\rm cos}^{-1} (a_{\rm min}/a_{\rm maj})$. The median values are $D=5.0$ kpc and $i=46^\circ$, slightly different from but still consistent with the typical values assumed in previous works \citep[e.g.,][]{Wang13,Willott15}. 
Note that when creating the moment zero maps of the [\ion{C}{2}] emission (as described in \S\ref{sec:obs}), we only collapsed the data cube within the frequency range of $\pm1.4\sigma_{\rm line}$. To assess whether this choice might miss any extended gas emission with higher velocities in the quasar host galaxies, we also measured the effective sizes of the [\ion{C}{2}] emitting regions by collapsing the data cube within the frequency range of $\pm3.0\sigma_{\rm line}$. From this experiment, we discovered that the ratio of the effective size derived from the wider cube width to that of the size derived from $\pm1.4\sigma_{\rm line}$ falls within the range of 0.86 to 1.27, with a median of $1.04\pm0.10$. This indicates the robustness of our size measurement. However, we acknowledge the possibility that we may have missed fainter emission at the outskirts of the galaxies due to limited data quality.

\subsection{Star Formation Rates}\label{subsec:sfr}
One of the main goals of this program is to measure the SFR of the targeted quasar host galaxies, which is important to our understanding of how the growth of the central SMBHs is related to the assembly of their host galaxies. In this work, we infer the SFR of quasar host galaxies using both [\ion{C}{2}] and dust continuum emissions. 

The [\ion{C}{2}] line is one of the primary fine-structure cooling lines and has been suggested as a good SFR tracer \citep[e.g.,][]{DeLooze14,HC15}. We use the empirical relation calibrated by \cite{DeLooze14} for high-redshift galaxies to estimate the SFR of these [\ion{C}{2}] detected quasar host galaxies:
\begin{equation}
\rm
log~SFR_{[CII]} [M_\odot ~ yr^{-1}] = -8.52 + 1.18 \times log~ L_{[CII]} [L_\odot].
\end{equation}

For the dust continuum-based SFR, we use two different templates as discussed above when measuring the $L_{\rm FIR}$: a modified graybody and the Haro11 galaxy template. 
In the case of the graybody in the optically regime, we assume a dust temperature of $T_{\rm dust}=47$ K and an emissivity index of $\beta=1.6$ \citep{Beelen06} and use the relation calibrated by \cite{Murphy11}:
\begin{equation}
\rm
SFR_{graybody} [M_\odot ~ yr^{-1}] = 3.88 \times 10^{-44}~ L_{TIR, graybody} [erg~s^{-1}].
\end{equation}
We use the star formation law calibrated by \cite{Lyu16} for the Haro 11 template:
\begin{equation}
\rm
SFR_{Haro11} [M_\odot ~ yr^{-1}] = 5.0 \times 10^{-44}~ L_{TIR,Haro11} [erg~s^{-1}].
\end{equation}

\begin{figure}
\centering
\includegraphics[width=0.48\textwidth]{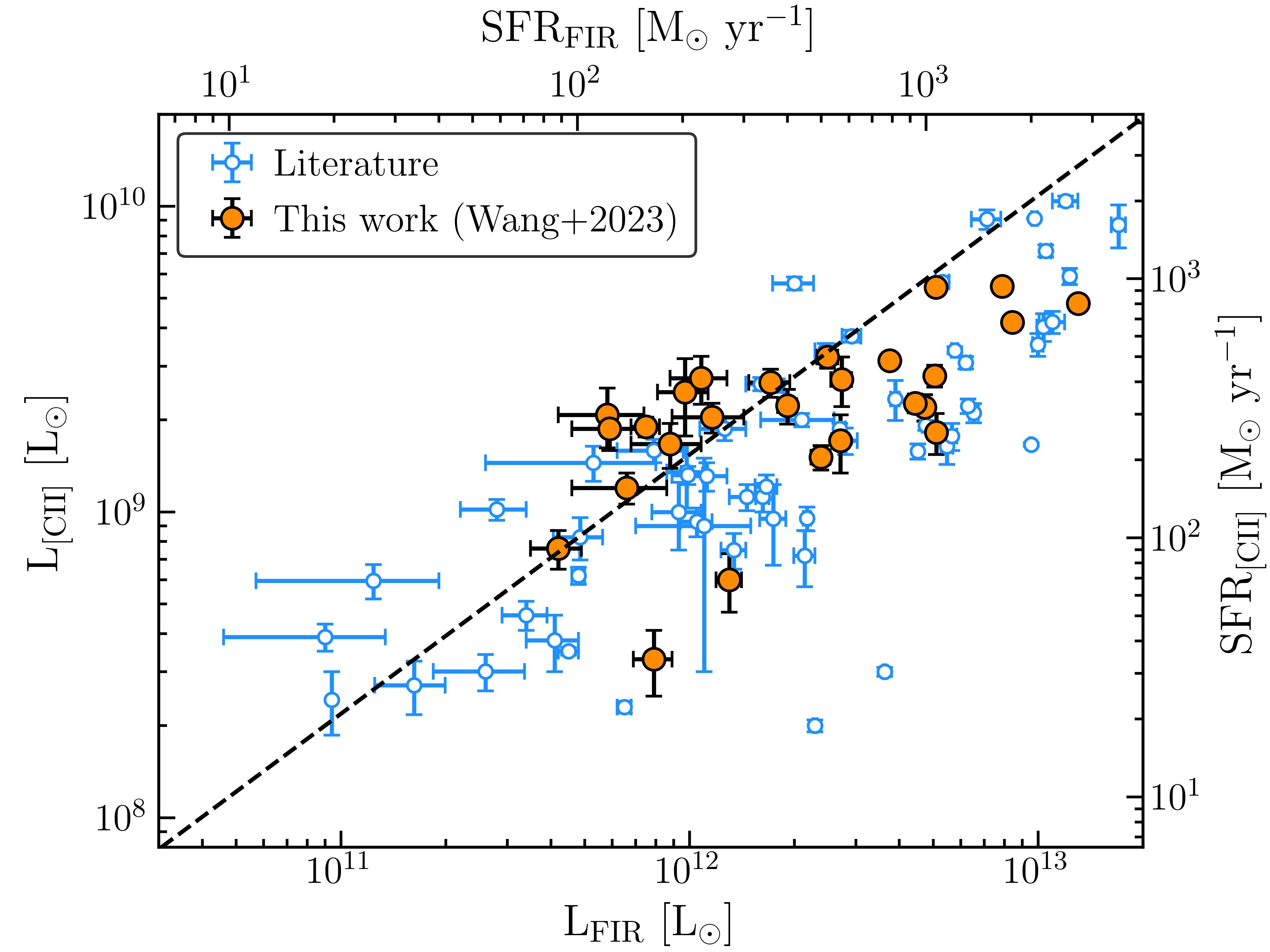}
\caption{ The relation between the [\ion{C}{2}] luminosity and the far-infrared luminosity. The star formation rates estimated from the [\ion{C}{2}] luminosity \citep{DeLooze14} and from the far-infrared luminosity \citep{Murphy11} are also shown. The black dashed line is the one-to-one case for the $\rm SFR_{[CII]}$ and the $\rm SFR_{FIR}$. 
\label{fig:c2fir}}
\end{figure}

\begin{deluxetable*}{ccccccccccccc}
\setlength{\tabcolsep}{2pt}
\tabletypesize{\scriptsize}
\tablecaption{Sizes of the [\ion{C}{2}] and continuum emission regions measured from 2D Gaussian fitting of ALMA data.}\label{tbl:size}
\tablehead{\colhead{Name} & \colhead{$R_{\rm [CII]}$} & \colhead{$A_{\rm [CII], ~Obs.}$} & \colhead{$A_{\rm [CII], ~Dec.}$} & \colhead{$A_{\rm [CII], ~Phy.}$} & \colhead{$\rm S/B_{\rm [CII]}$} & 
                  \colhead{$R_{\rm cont}$} & \colhead{$A_{\rm cont,~Obs.}$} &  \colhead{$A_{\rm cont, ~Dec.}$} & \colhead{$A_{\rm cont, ~Phy.}$} & \colhead{$\rm S/B_{\rm cont}$} }
\startdata
        &   [kpc]  &  [$^{\prime\prime}\times^{\prime\prime}$]  &  [$^{\prime\prime}\times^{\prime\prime}$]  &  [$\rm kpc \times kpc$] & &   [kpc]  &  [$^{\prime\prime}\times^{\prime\prime}$]  &  [$\rm kpc \times kpc$] \\
\hline
J0038$-$1527 & 2.45$\pm$0.23 & 1.12$\times$0.97 & 0.94$\times$0.80 & 4.90$\times$4.17 & 1.87$\pm$0.12 & 1.56$\pm$0.18 & 0.86$\times$0.68 & 0.60$\times$0.38 & 3.13$\times$1.98 & 1.39$\pm$0.06 \\
J0213$-$0626 & -- & -- & -- & -- & -- & 1.15$\pm$0.34 & 0.74$\times$0.59 & 0.44$\times$0.15 & 2.29$\times$0.78 & 1.39$\pm$0.06 \\
J0218$+$0007 & 1.95$\pm$0.48 & 0.83$\times$0.68 & 0.73$\times$0.58 & 3.89$\times$3.09 & 2.18$\pm$0.39 & 0.83$\pm$0.11 & 0.50$\times$0.47 & 0.31$\times$0.29 & 1.65$\times$1.55 & 1.28$\pm$0.05 \\
J0224$-$4711 & 1.42$\pm$0.08 & 0.76$\times$0.75 & 0.52$\times$0.48 & 2.83$\times$2.62 & 1.33$\pm$0.04 & 1.01$\pm$0.05 & 0.69$\times$0.68 & 0.37$\times$0.36 & 2.02$\times$1.96 & 1.19$\pm$0.02 \\
J0229$-$0808 & 1.20$\pm$0.08 & 0.60$\times$0.53 & 0.45$\times$0.41 & 2.41$\times$2.19 & 1.54$\pm$0.05 & 0.62$\pm$0.03 & 0.46$\times$0.41 & 0.23$\times$0.20 & 1.23$\times$1.07 & 1.15$\pm$0.00 \\
J0246$-$5219 & 1.35$\pm$0.11 & 0.64$\times$0.56 & 0.51$\times$0.44 & 2.69$\times$2.32 & 1.68$\pm$0.08 & 0.74$\pm$0.05 & 0.48$\times$0.44 & 0.28$\times$0.25 & 1.48$\times$1.32 & 1.23$\pm$0.03 \\
J0252$-$0503 & 3.45$\pm$0.71 & 1.39$\times$0.82 & 1.32$\times$0.69 & 6.90$\times$3.61 & 2.90$\pm$0.50 & 1.07$\pm$0.16 & 0.64$\times$0.50 & 0.41$\times$0.30 & 2.14$\times$1.57 & 1.28$\pm$0.06 \\
J0313$-$1806 & 1.36$\pm$0.55 & 0.78$\times$0.60 & 0.55$\times$0.40 & 2.73$\times$1.98 & 1.32$\pm$0.22 & 1.12$\pm$0.22 & 0.72$\times$0.53 & 0.45$\times$0.28 & 2.23$\times$1.39 & 1.24$\pm$0.09 \\
J0319$-$1008 & 2.52$\pm$0.77 & 1.03$\times$0.79 & 0.95$\times$0.72 & 5.04$\times$3.82 & 2.64$\pm$0.69 & 0.53$\pm$0.29 & 0.44$\times$0.38 & 0.20$\times$0.16 & 1.06$\times$0.85 & 1.10$\pm$0.10 \\
J0411$-$0907 & 4.91$\pm$1.19 & 1.95$\times$1.53 & 1.85$\times$1.41 & 9.82$\times$7.48 & 2.87$\pm$0.59 & 1.65$\pm$0.74 & 0.95$\times$0.66 & 0.62$\times$0.34 & 3.29$\times$1.80 & 1.32$\pm$0.26 \\
J0430$-$1445 & 1.74$\pm$0.21 & 0.74$\times$0.56 & 0.65$\times$0.46 & 3.48$\times$2.46 & 2.00$\pm$0.19 & 1.42$\pm$0.27 & 0.64$\times$0.48 & 0.53$\times$0.34 & 2.84$\times$1.82 & 1.68$\pm$0.18 \\
J0525$-$2406 & 1.71$\pm$0.05 & 0.72$\times$0.50 & 0.63$\times$0.35 & 3.43$\times$1.90 & 2.00$\pm$0.06 & 0.84$\pm$0.03 & 0.47$\times$0.41 & 0.31$\times$0.20 & 1.69$\times$1.09 & 1.27$\pm$0.03 \\
J0706$+$2921 & 2.98$\pm$0.62 & 1.40$\times$1.06 & 1.10$\times$0.81 & 5.95$\times$4.38 & 1.56$\pm$0.17 & 1.79$\pm$0.30 & 1.12$\times$0.74 & 0.66$\times$0.31 & 3.57$\times$1.68 & 1.22$\pm$0.07 \\
J0910$+$1656 & 3.24$\pm$0.54 & 1.34$\times$1.12 & 1.21$\times$0.89 & 6.48$\times$4.76 & 1.97$\pm$0.24 & 1.61$\pm$0.91 & 0.88$\times$0.70 & 0.60$\times$0.31 & 3.21$\times$1.66 & 1.26$\pm$0.21 \\
J0910$-$0414 & 1.48$\pm$0.19 & 0.81$\times$0.78 & 0.55$\times$0.52 & 2.97$\times$2.81 & 1.31$\pm$0.05 & 0.62$\pm$0.05 & 0.68$\times$0.61 & 0.23$\times$0.23 & 1.24$\times$1.24 & 1.06$\pm$0.00 \\
J0921$+$0007 & 3.09$\pm$0.43 & 1.29$\times$0.89 & 1.14$\times$0.70 & 6.19$\times$3.80 & 2.15$\pm$0.23 & 3.77$\pm$1.30 & 1.51$\times$0.79 & 1.39$\times$0.51 & 7.55$\times$2.77 & 2.44$\pm$0.68 \\
J0923$+$0402 & 2.46$\pm$0.32 & 1.07$\times$1.05 & 0.91$\times$0.87 & 4.91$\times$4.70 & 1.78$\pm$0.17 & 3.62$\pm$0.92 & 1.47$\times$0.98 & 1.34$\times$0.79 & 7.23$\times$4.26 & 2.41$\pm$0.49 \\
J0923$+$0753 & 3.04$\pm$0.54 & 1.19$\times$0.91 & 1.13$\times$0.82 & 6.07$\times$4.41 & 3.05$\pm$0.46 & 2.15$\pm$0.51 & 0.89$\times$0.56 & 0.80$\times$0.44 & 4.30$\times$2.36 & 2.23$\pm$0.43 \\
J1007$+$2115 & 1.10$\pm$0.20 & 0.62$\times$0.53 & 0.44$\times$0.34 & 2.20$\times$1.70 & 1.32$\pm$0.09 & 1.08$\pm$0.08 & 0.62$\times$0.48 & 0.43$\times$0.32 & 2.15$\times$1.60 & 1.35$\pm$0.04 \\
J1058$+$2930 & 2.71$\pm$0.68 & 1.08$\times$0.95 & 1.00$\times$0.82 & 5.42$\times$4.44 & 2.08$\pm$0.33 & 1.44$\pm$0.51 & 0.75$\times$0.37 & 0.53$\times$0.14 & 2.87$\times$0.76 & 1.42$\pm$0.23 \\
J1104$+$2134 & 2.13$\pm$0.37 & 0.99$\times$0.97 & 0.80$\times$0.68 & 4.27$\times$3.63 & 1.41$\pm$0.16 & 0.53$\pm$0.13 & 0.75$\times$0.63 & 0.20$\times$0.16 & 1.07$\times$0.85 & 1.04$\pm$0.01 \\
J1129$+$1846 & -- & -- & -- & -- & -- & -- & -- & -- & -- & -- \\
J2002$-$3013 & 1.21$\pm$0.19 & 0.57$\times$0.49 & 0.45$\times$0.35 & 2.42$\times$1.88 & 1.50$\pm$0.11 & 0.48$\pm$0.05 & 0.41$\times$0.35 & 0.18$\times$0.07 & 0.97$\times$0.38 & 1.08$\pm$0.03 \\
J2102$-$1458 & 1.80$\pm$0.27 & 0.88$\times$0.68 & 0.67$\times$0.45 & 3.61$\times$2.42 & 1.49$\pm$0.10 & 1.13$\pm$0.22 & 0.72$\times$0.64 & 0.42$\times$0.35 & 2.26$\times$1.88 & 1.18$\pm$0.05 \\
J2211$-$6320 & 1.25$\pm$0.16 & 0.71$\times$0.51 & 0.47$\times$0.26 & 2.49$\times$1.38 & 1.37$\pm$0.08 & 0.58$\pm$0.08 & 0.59$\times$0.51 & 0.22$\times$0.16 & 1.17$\times$0.85 & 1.05$\pm$0.02
\enddata
\end{deluxetable*}

\begin{figure*}
\centering
\includegraphics[width=0.98\textwidth]{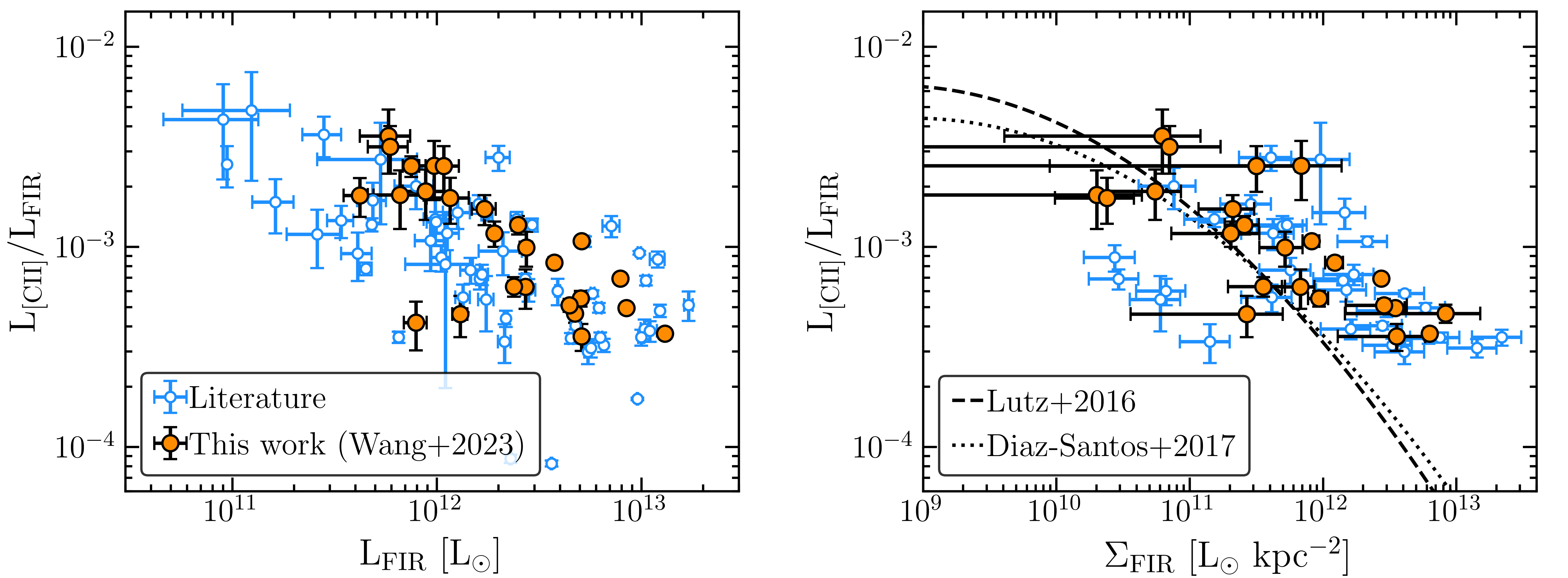}
\caption{ The [\ion{C}{2}]-to-FIR luminosity ratio as a function of FIR luminosity (\emph{left}) and  the FIR luminosity surface density, $\Sigma_{\rm FIR}$ (\emph{right}).
The [\ion{C}{2}]-to-FIR luminosity ratios of these high redshift quasars span a wide range (nearly two orders of magnitude). In the right panel, we also plot the best fits found by \cite{Lutz16} and \cite{Diaz-Santos17} from local-infrared luminous galaxies.
The [\ion{C}{2}] deficit is clearly seen in these plots but the relations have substantial scatter. 
\label{fig:deficit}}
\end{figure*}

The SFRs derived from all three estimators are listed in Table \ref{tbl:measure}. 
In Figure \ref{fig:c2fir}, we show the relation between the SFR derived from the [\ion{C}{2}] lines and the dust continuum-based SFR inferred from the graybody model. The SFRs derived from these two methods are broadly consistent with each other, with a range of $\rm SFR\sim 100-3000~M_\odot ~ yr^{-1}$ and both are about two times lower than the SFR derived from the Haro 11 template. 
However, the [\ion{C}{2}] based SFR at the high-luminosity end is systematically lower than that derived from the FIR-based SFR. This is likely due to the fact that the empirical relation from \cite{DeLooze14} was calibrated to star-forming galaxies, while the FIR-luminous galaxies have a significant [\ion{C}{2}] deficit effect at the high-luminosity end. We will discuss this in \S \ref{subsec:deficit} in more detail.
Since the $\rm SFR_{graybody}$ is more broadly used in the literature and more consistent with the $\rm SFR_{[CII]}$, we will only use the $\rm SFR_{graybody}$ and $\rm SFR_{[CII]}$ in the following analyses. 
We note that the SFRs derived from all three methods have large uncertainties given that we have no knowledge about the metallicity and dust temperature of the quasar host galaxies. 
In particular, the scaling relations were derived from nominal star-forming galaxies ($L_{\rm [CII]}$ and $L_{\rm TIR, graybody}$) or local galaxies (Haro 11) which have not been validated as accurate templates for high-redshift quasar host galaxies.
Such uncertainties can be better understood with future ALMA band 8/9/10 and JCMT/SCUBA-2 high frequency observations and JWST mid-infrared observations by capturing the peak of the dust continuum emission and improving our understanding of the AGN contribution to dust heating, respectively.
The inferred SFR surface densities (i.e., the graybody-based SFR divided by our estimates of the continuum emitting size) are in the range of $\sim$1--800 $\rm M_\odot~yr^{-1}~kpc^{-1}$, below the star formation Eddington limit \citep[$\sim1000~\rm M_\odot~yr^{-1}~kpc^{-1}$;][]{Walter09, Walter22}.

\subsection{Dust and Gas Mass} \label{subsec:dust_mass}
The dust masses of quasar host galaxies can be estimated from the observed FIR flux density with an assumption of the dust temperature, emissivity index with an optically thin case at the observed wavelength using the following equation:
\begin{equation}
M_{\rm dust} = \frac{F_\nu D_L^2}{(1+z) \kappa_\nu(\beta) [B_\nu(\nu, T)-B_\nu(\nu, T_{\rm CMB})]},
\label{eq:dust}
\end{equation}
where $F_\nu$ is the continuum flux density measured from the 2D continuum map as listed in Table \ref{tbl:c2fit}, $D_L$ is the luminosity distance, $\kappa_\nu(\beta)$ is the dust mass opacity coefficient, and $B_\nu$ is the Planck function. Following \cite{Venemans18}, the opacity coefficient is set by $\kappa_\nu(\beta)=0.77\,(\nu/352\,\mathrm{GHz})^\beta$\,cm$^2$\,g$^{-1}$. To be consistent with our measurements of $L_{\rm FIR}$ and the reports from most of the literature, we assume $T=47$ K and $\beta=1.6$ in Eq. \ref{eq:dust}, and take the CMB effect, $B_\nu(\nu, T_{\rm CMB})$, into account when estimating the dust mass.
The estimated dust masses of these continuum-detected quasar host galaxies range from $\sim3\times10^7~M_\odot$ to $\sim8\times10^8~M_\odot$ and are listed in Table \ref{tbl:measure}. The dust contents of these quasar host galaxies are similar to those found in $z\sim6$ quasar host galaxies \citep[e.g.][]{Izumi18,Venemans18}. 
By assuming a dust-to-gas ratio of 100 and a molecular-to-total gas mass fraction of 0.75 \citep[e.g.][]{Neeleman21}, we can roughly estimate the gas mass in these galaxies to be in the range of $4\times10^9~M_\odot$ to $1\times10^{11}~M_\odot$, suggesting that these quasar host galaxies are among the most massive and gas-rich galaxies in the early universe. 
However, we note that the dust mass inferred from a single continuum measurement has large uncertainties and could increase by a factor of two if we assume a lower dust temperature (i.e. $T=35$ K). Future high-frequency FIR continuum observations and CO line observations are needed to improve the constraints on the dust and gas masses. 

The mass of singly ionized carbon can be derived from the strength of the [\ion{C}{2}] emission line. Following \cite{Venemans17b}, we calculate this quantity using the following equation by assuming that the emission is optically thin \cite{Weiss05}. 
\begin{equation}
M_{\rm {C}^+} =  2.92\times10^{-4} Q(T_{\rm ex}) \frac{1}{4} {\rm e}^{91.2/T_{\rm ex}} 
L^\prime_{\rm [CII]},
\label{eq:c2mass}
\end{equation}
where $T_{\rm ex}$ is the excitation temperature, and $Q(T_{\rm ex})=2+4{\rm e}^{-91.2/T_{\rm ex}}$ is the [\ion{C}{2}] partition function.
By assuming $T_{\rm ex}=100$ K \citep[e.g.][]{Meijerink07}, we derive the mass of singly ionized carbon from the measured [\ion{C}{2}] luminosity, $L^\prime_{\rm [CII]}$ (in units of K km s$^{-1}$ pc$^2$).
The measured $M_{\rm {C}^+}$ has a broad range between $8\times10^5~M_\odot$ and $1.6\times10^7~M_\odot$. The $M_{\rm {C}^+}$ values for [\ion{C}{2}]-detected galaxies are also given in Table \ref{tbl:measure}.

\begin{figure}
\centering
\includegraphics[width=0.49\textwidth]{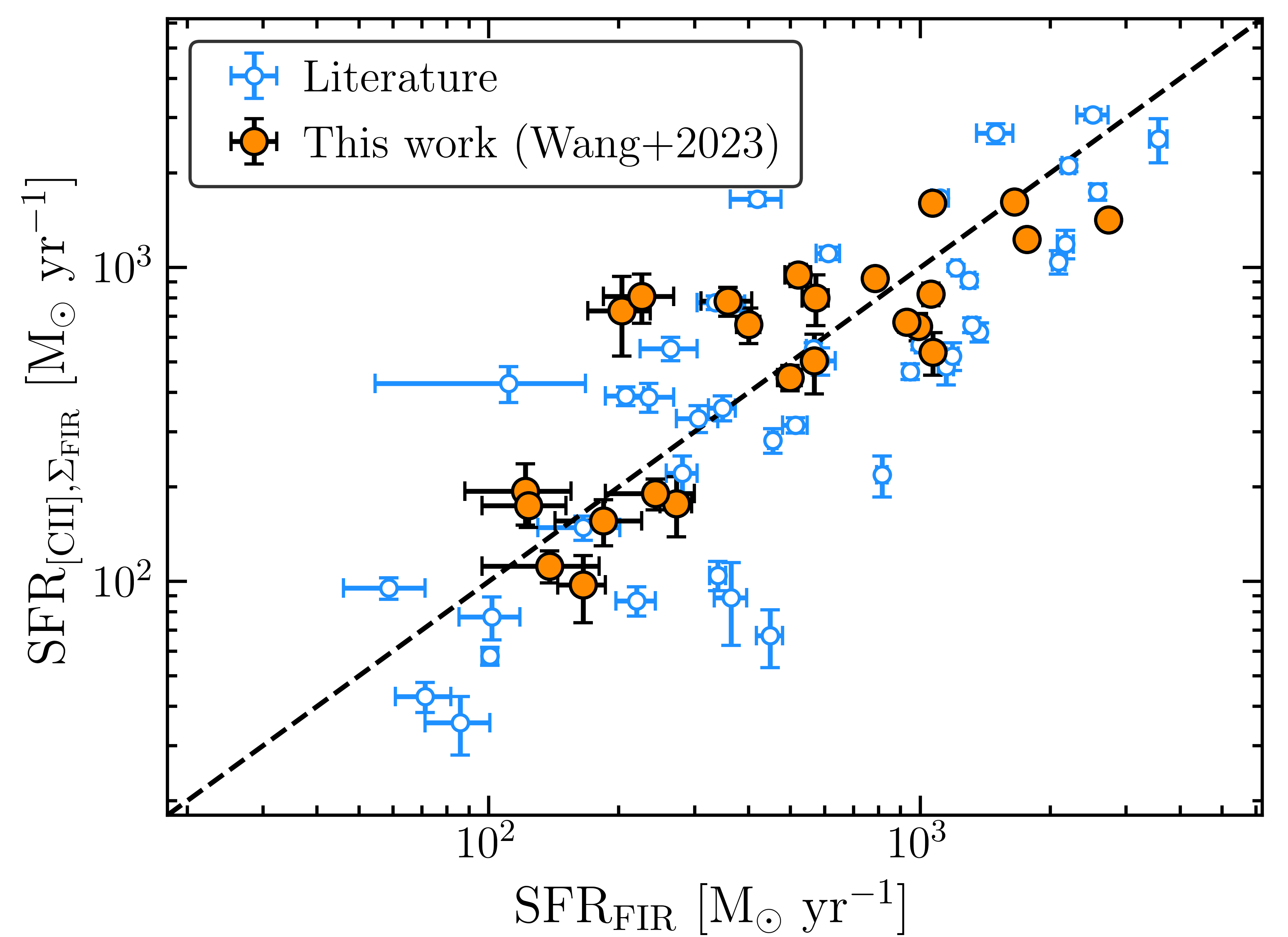}
\caption{
The [\ion{C}{2}] based SFR is consistent with FIR based SFR after considering the FIR surface brightness of the quasar host galaxies. The x-axis in this plot is the SFR estimated from the $L_{\rm FIR}$, same as Figure \ref{fig:Lc2}. The y-axis is the SFR estimated from $L_{\rm [CII]}$ using the relations from \cite{HC18} after considering the FIR surface brightness of the quasar host galaxies. 
\label{fig:sfr_corr}}
\end{figure}

\begin{figure*}
\centering
\includegraphics[width=0.98\textwidth]{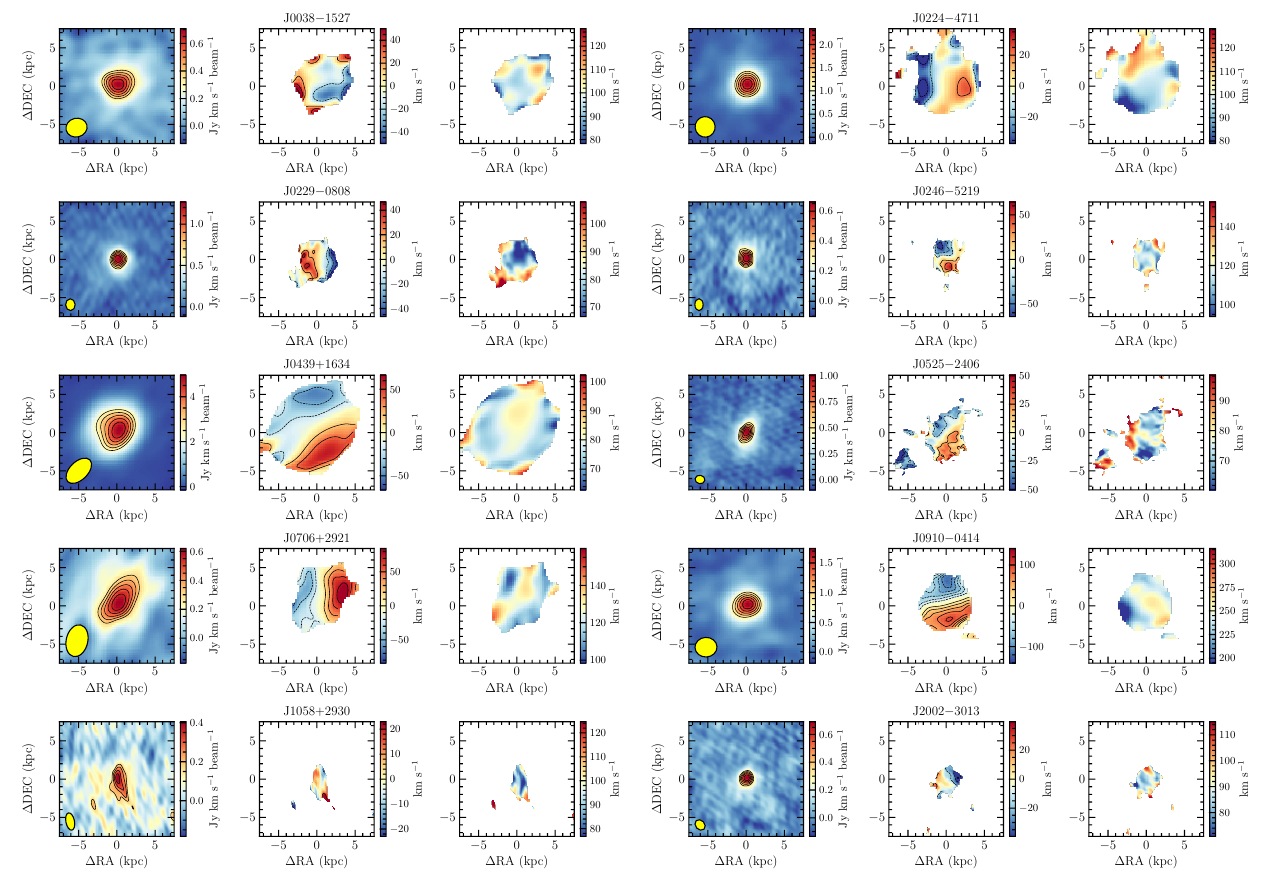}
\vspace{-5pt}
\caption{High-resolution ALMA moment 0 (left), 1 (middle), and 2 (right) maps of 12 objects with clear [\ion{C}{2}] velocity gradients. Each panel is $15 \times 15 ~ {\rm kpc^2}$ wide. North is up and east to the left. In the moment 0 map, the solid black contours mark the +3, 4, 5, 6 isophotes for those objects with continuum peak at $<7\sigma$, or the 50\%, 60\%, 70\%, 80\%, 90\% of the continuum peak emission for those objects with $\ge7\sigma$  peak detections. The synthesized beam of the observations is shown in the bottom-left corner of each panel. 
The solid contours and dashed contours in the moment 1 maps mark the positive velocities and negative velocities, respectively. The contour levels for both positive velocities and negative velocities are in steps of 20 km s$^{-1}$. 
}\label{fig:mom_vel}
\end{figure*}

\begin{figure*}
\centering
\includegraphics[width=0.98\textwidth]{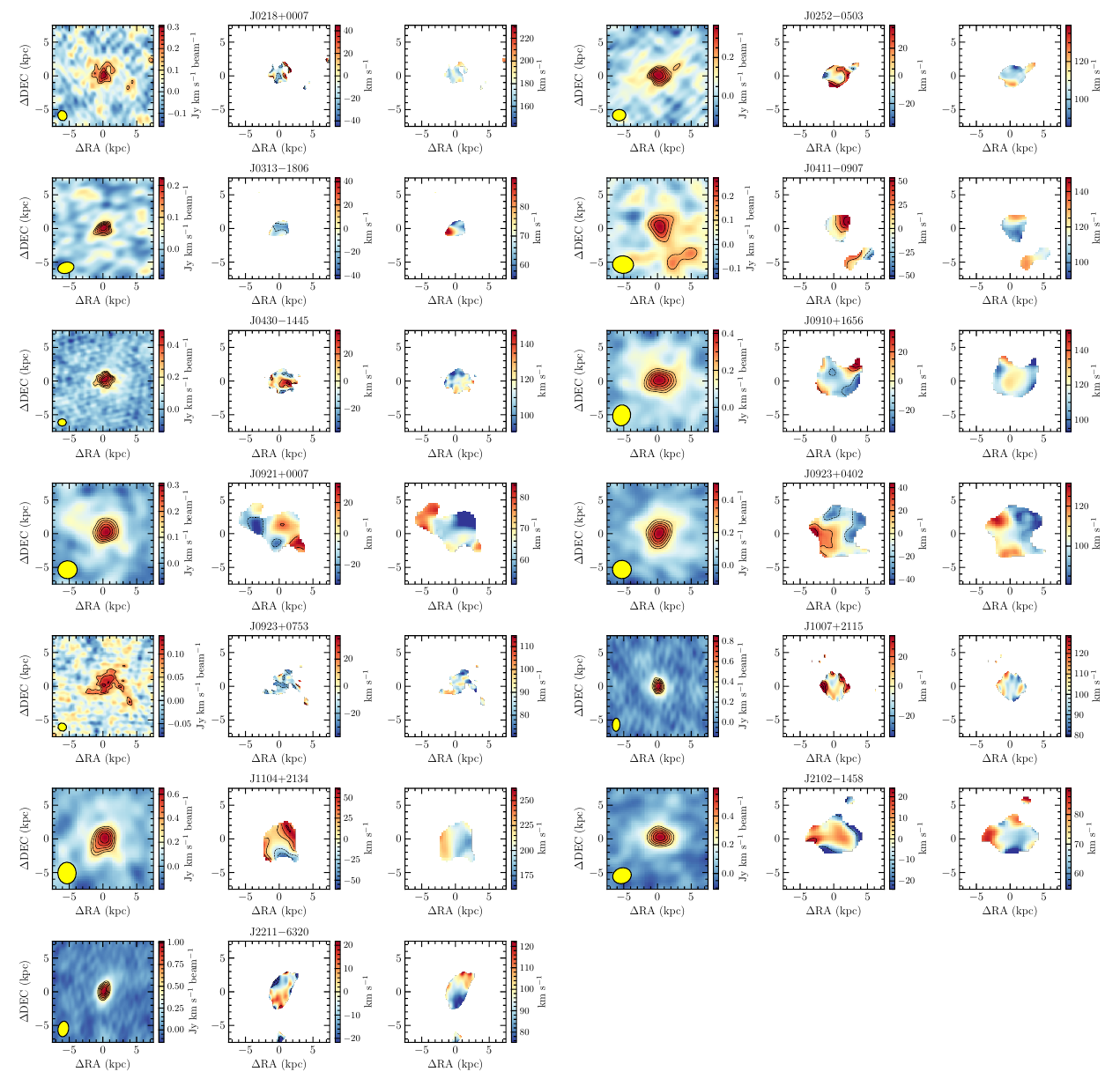}
\vspace{-5pt}
\caption{Same as Figure \ref{fig:mom_vel}, but for 11 galaxies without clear velocity gradients. 
}\label{fig:mom_other}
\end{figure*}

\subsection{[\ion{C}{2}] Deficit} \label{subsec:deficit}
The ratio between $L_{\rm [CII]}$ and $L_{\rm FIR}$ is an indicator of the contribution of the [\ion{C}{2}] line to the cooling of the interstellar medium. The $L_{\rm [CII]}$/$L_{\rm FIR}$ was found to be $\sim10^{-3}$ for Milky Way-like star-forming galaxies, while it is about an order of magnitude smaller in FIR-luminous systems like ULIRGs \citep[see][for a recent review]{Hodge20}. Although the underlying physical causes of the observed `[\ion{C}{2}] deficit' are still debated, subsequent studies have expanded the investigation of the [\ion{C}{2}] deficit to high-redshift galaxies and to sub-kpc scales, indicating that the physical scale of [\ion{C}{2}] deficits must be sub-kpc \citep[e.g.][]{Lamarche18} and could be caused by a locally intense radiation field \citep[e.g.][]{Rybak19}. 

In Figure \ref{fig:deficit}, we show $L_{\rm [CII]}$/$L_{\rm FIR}$ as a function of both $L_{\rm FIR}$ and the FIR luminosity surface density, $\Sigma_{\rm FIR}$. We compute 
$\Sigma_{\rm FIR}=L_\mathrm{FIR} / (2\pi R^2_\mathrm{eff,cont})$
based on the 2D Gaussian fit of the continuum map as detailed in \S \ref{subsec:size}, where $R_{\rm eff, cont}=\sqrt{R_{\rm maj, cont}R_{\rm min, cont}}$. Following \cite{Decarli18}, we include a factor 2 in the denominator when computing $\Sigma_{\rm FIR}$ to account for the fact that $R_{\rm eff, cont}$ roughly encompasses half of the total light.
The $L_{\rm [CII]}$/$L_{\rm FIR}$ ratios of these high-redshift quasar host galaxies span nearly two orders of magnitude. In the right panel of Figure \ref{fig:deficit}, we also show the [\ion{C}{2}] deficit relations found from local infrared-luminous galaxies.
The relation between $L_{\rm [CII]}$/$L_{\rm FIR}$ and $\Sigma_{\rm FIR}$ in quasar host galaxies is similar to that of local infrared-luminous galaxies, indicating that the [\ion{C}{2}] deficit is not strongly affected by the presence of luminous AGN \citep[see also][]{Venemans20}. 
However, we need to acknowledge that the current dataset does not allow us to understand the physical origins of the  [\ion{C}{2}] deficit. 
In this work, we assumed $T=47$ K for all quasar host galaxies which could introduce significant uncertainties on the $L_{\rm FIR}$ measurements. In addition, the current observations only marginally resolve the quasar host galaxies and do not allow us to study $\Sigma_{\rm FIR}$ as a function of location. Furthermore, we do not have metallicity constraints on these quasar host galaxies. 
Future multiple-frequency continuum observations, high-resolution observations of both dust continuum and [\ion{C}{2}] and multi-line diagnostics are necessary to constrain the  physical origins of the [\ion{C}{2}] deficit.

As discussed in \S \ref{subsec:sfr}, the [\ion{C}{2}]-based SFR derived from \cite{DeLooze14} is systematically lower than that derived from $L_{\rm FIR}$ at the high-luminosity end. 
To test whether the underestimated [\ion{C}{2}]-based SFR is caused by using the nominal star-forming galaxy empirical relation, 
we computed the [\ion{C}{2}]-based SFR using the relations from \cite{HC18} for sources with $\Sigma_{\rm SFR}$ measurements.
\cite{HC18} studied the [\ion{C}{2}]–SFR scaling relations in different types of galaxies and found that the ratio of $L_{\rm [CII]}$/SFR is different for galaxies with different $\Sigma_{\rm SFR}$, consistent with the [\ion{C}{2}] deficit observed in these galaxies: 
$\rm log_{10}(L_{\rm [CII]}/SFR)=7.03\pm0.21$ for galaxies with $\Sigma_{\rm SFR}$ in the range of $\rm 10^8-10^{11.2}~L_\odot~kpc^{-2}$ and $\rm log_{10}(L_{\rm [CII]}/SFR)=6.53\pm0.30$ for galaxies with $\Sigma_{\rm SFR}$ in the range of $\rm 10^{11.2}-10^{12}~L_\odot~kpc^{-2}$. 
The comparison of these measurements with the $L_{\rm FIR}$ based SFR is shown in Figure \ref{fig:sfr_corr}. This figure indicates that the [\ion{C}{2}]-based SFR after considering the $\Sigma_{\rm SFR}$ is more consistent with the $L_{\rm FIR}$-based SFR when compared with Figure \ref{fig:c2fir}. This test suggests that the [\ion{C}{2}] deficit in quasar host galaxies is broadly consistent with star-forming galaxies at lower redshifts (see also Figure \ref{fig:deficit}) and it is necessary to consider the FIR radiation field intensities when estimating the SFR of quasar host galaxies using the [\ion{C}{2}] line.

\begin{figure*}
\centering
\includegraphics[width=0.98\textwidth]{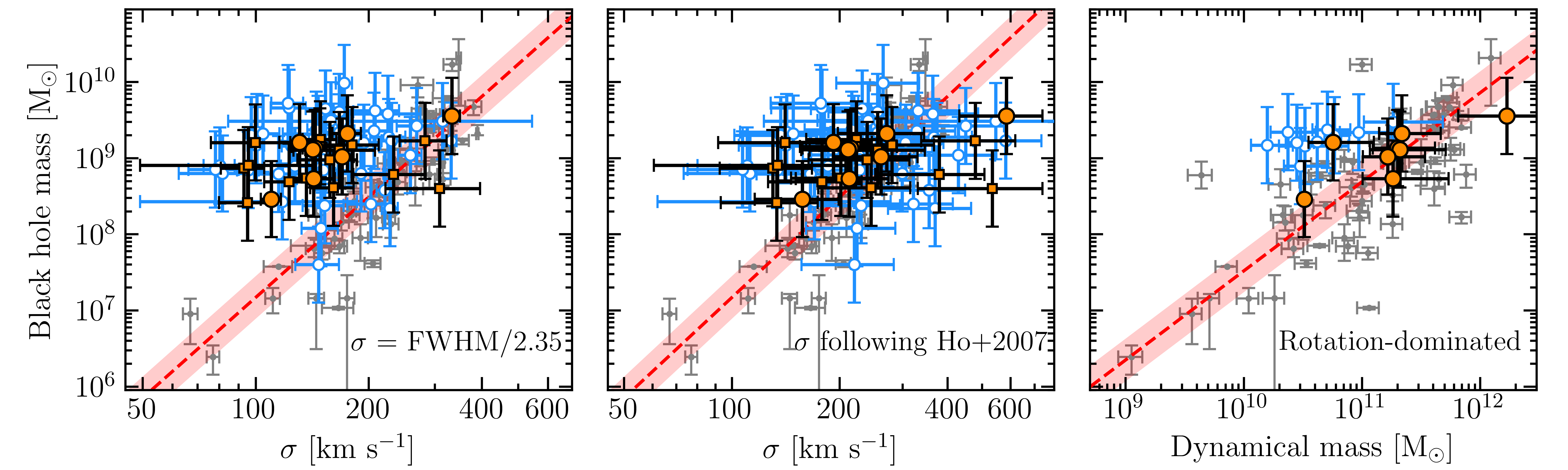}
\caption{Early universe SMBH-host galaxy co-evolution. In all panels, the orange circles and orange squares denote our measurements of $z>6.5$ quasars with rotation-dominated kinematics and dispersion-dominated kinematics or disturbed galaxies, respectively. The open blue points represent data of $z\sim6$ quasars collected from literature, and the gray points are the measurements of local galaxies from \cite{KH13}. The best fits for the local $M_{\rm BH}-\sigma$ and $M_{\rm BH}-M_{\rm bulge}$ relations derived by \cite{KH13} are shown as a red dashed line with the shaded regions denoting the 1$\sigma$ scatter of $M_{\rm BH}$ and $\sigma$/$M_{\rm bulge}$. The error bars of $M_{\rm BH}$ in this plot include the uncertainty from the \ion{Mg}{2} black hole mass scaling relation, which is $\sim0.5$ dex \citep{Vestergaard09}. 
{\it Left:} Black hole mass versus $\sigma$ derived from $\rm FWHM_{[CII]}$.
{\it Middle:} The black hole mass versus $\sigma$ derived using the method proposed by \cite{Ho07} which corrects for the effects of inclination and gas turbulence. The uncertainties of $\sigma_{\rm Ho}$ includes both measurement uncertainty and the relation uncertainty in \cite{Ho07}.
{\it Right:} The black hole mass versus dynamical mass estimated from Equation \ref{eq:dynamical}. 
\label{fig:msigma}}
\end{figure*}

\section{Kinematics and Dynamical Mass} \label{sec:mass}
\subsection{Kinematics and Morphology} \label{subsec:morphology}
\cite{Neeleman21} presented kinematic analyses of 27 quasar host galaxies at $z\sim6$ with ALMA at $\sim$0\farcs1--0\farcs25 resolution.
They found that about one-third of the $z\sim6$ quasar host galaxies (10 out of 27) show a smooth velocity gradient, one-third (9 out of 27) of the quasar host galaxies have disturbed [\ion{C}{2}] emission profiles, and the remaining one-third (8 out of 27) show a velocity profile without a clear velocity gradient, which is consistent with the emission arising from a dispersion-dominated system. They also found that all galaxies have intrinsically large velocity dispersions despite different kinematics. 

To investigate the kinematics of the quasar host galaxies from our $z\sim7$ quasar sample, we re-image the ALMA [\ion{C}{2}] data cube using the {\tt tclean} task with robust parameter $r=0.5$ to obtain images with slightly higher spatial resolution ($\sim$0\farcs2--0\farcs5). Three objects observed with ALMA (J0213--0626, J0319--1008, and J1129+1846) have relatively low significance detections, hence we only focus on the 23 objects observed with ALMA. 
The [\ion{C}{2}] intensity map is collapsed from $-1.4\sigma_{\rm line}$ to $+1.4\sigma_{\rm line}$ of the emission-line data cube. 
We then create a mask image by selecting all pixels with S/N$>3$ in the intensity map. We finally create a velocity map (moment 1) and a velocity dispersion map (moment 2) of each target using the mask image created from previous step and selecting frequencies between $-2.0\sigma_{\rm line}$ and $+2.0\sigma_{\rm line}$ with the {\tt immoments} task within CASA package. The moment maps are shown in Figure \ref{fig:mom_vel} and Figure \ref{fig:mom_other}. Similar to what has been found by \cite{Neeleman21}, all galaxies show large velocity dispersions as indicated by the moment 2 maps, although the marginally resolved maps are affected by beam smearing and by the uncertainty of the inclination angles. 
\cite{Forster09} proposed a general method for distinguishing rotation-dominated galaxies from dispersion-dominated systems by using the ratio of the projected velocity gradient ($v_{\rm proj}$, observed difference between the maximum and minimum relative velocities without correcting for inclination) and the integrated line width ($\sigma_{\rm int}$, measured from 1D line profile) for marginally resolved observations. Based on simulations of disk galaxies, they found that galaxies with $v_{\rm proj}/2\sigma_{\rm int}>0.4$ can be classified as rotation-dominated systems. Such method has been successfully used in ALMA observations of high-redshift galaxies \citep{Smit18}. We measure $v_{\rm proj}$ from the velocity map shown in Figure \ref{fig:mom_vel} and Figure \ref{fig:mom_other} and $\sigma_{\rm int}$ from the measured line widths in Table \ref{tbl:c2fit}. The measured $v_{\rm proj}/2\sigma_{\rm int}$ for the 23 quasar host galaxies are listed in Table \ref{tbl:c2fit}. Nine of the 23 galaxies have $v_{\rm proj}/2\sigma_{\rm int}>0.4$, however, J0923+0753 shows obvious distorted morphology. Further more, two galaxies have $v_{\rm proj}/2\sigma_{\rm int}<0.4$ but with clear velocity gradient (J0246--5219 and J0910--0414). We therefore conclude to classify  J0038--1527, J0224--4711, J0229--0808, J0246--5219, J0439+1634, J0525--2406, J0706+2921, J0910--0414, J1058+2930, and J2002--3013 (43\%) as rotation-dominated galaxies (Fig. \ref{fig:mom_vel}). The remaining galaxies (57\%) are classified are likely either dispersion-dominated galaxies or disturbed by galaxy mergers. Such fraction is similar to that found in the $z\sim6$ quasar host galaxies in \cite{Neeleman21} if we also separate their galaxies into dispersion-dominated kinematics (13 out of 27) and galaxies showing clear velocity gradients (14 out of 27) solely based on velocity maps. We note that such method could cause some mis-classification limited by the low-resolution of the data used in this work and future high-resolution observations are needed to improve our classifications.

\subsection{Dynamical Mass} \label{subsec:dynamical}
In the following, we estimate the dynamical mass for those rotation-dominated galaxies (except for the lensed galaxy J0439+1634) by assuming that the [\ion{C}{2}] emission is originating from a thin rotating disk.
As mentioned in \S \ref{subsec:size}, the disk size, $D$, is given by 1.5$\times$ the FWHM major axis from the [\ion{C}{2}] intensity map, and the factor 1.5 is used to account for spatially extended low-level emission \citep[e.g.][]{Wang16}. The inclination angle, $i$, is given by the ratio of minor ($a_{\rm min}$) and major ($a_{\rm maj}$) axes, $i = {\rm cos}^{-1} (a_{\rm min}/a_{\rm maj})$. The circular velocity is expressed as $v_{\rm circ} = 0.75\times {\rm FWHM_{[C II]}} / {\rm sin}~i$. The dynamical mass within $D$ can then be estimated by:
\begin{equation}
M_{\rm dyn}~[M_\odot] =  1.16\times10^{5} \left(\frac{v_{\rm circ}}{\rm km~s^{-1}}\right)^2 \left(\frac{D}{\rm kpc}\right).
\label{eq:dynamical}
\end{equation}
We use the size and the inclination angle measured in \S \ref{subsec:size}.
The dynamical masses of these galaxies estimated using Equation \ref{eq:dynamical} are listed in Table \ref{tbl:measure}.

\cite{Neeleman21} note that the dynamical mass measured from Equation \ref{eq:dynamical} using marginally resolved observations could be overestimated because of the presence of high velocity dispersion. They suggested an empirical relation based on their high-resolution [\ion{C}{2}] observations:
\begin{equation}
M_{\rm dyn}~[M_\odot] =  5.8\times10^{4} \left(\frac{{\rm FWHM}/{{\rm sin}~i}}{\rm km~s^{-1}}\right)^2 \left(\frac{R}{\rm kpc}\right).
\label{eq:dynamical_N21}
\end{equation}
The dynamical masses measured using Equation \ref{eq:dynamical_N21}, which are $\sim2.3$ times lower than those estimated from Equation \ref{eq:dynamical}, are also listed in Table \ref{tbl:measure}. The dynamical mass spans a broad range of $\sim(0.1-7.5)\times10^{11}~M_\odot$. We emphasize that the $M_{\rm dyn}$ estimated from both equations have large uncertainties, we will improve the $M_{\rm dyn}$ measurements based on kinematic modeling of future high-resolution ALMA observations (Yang et al. in prep).

\begin{figure}
\centering
\includegraphics[width=0.48\textwidth]{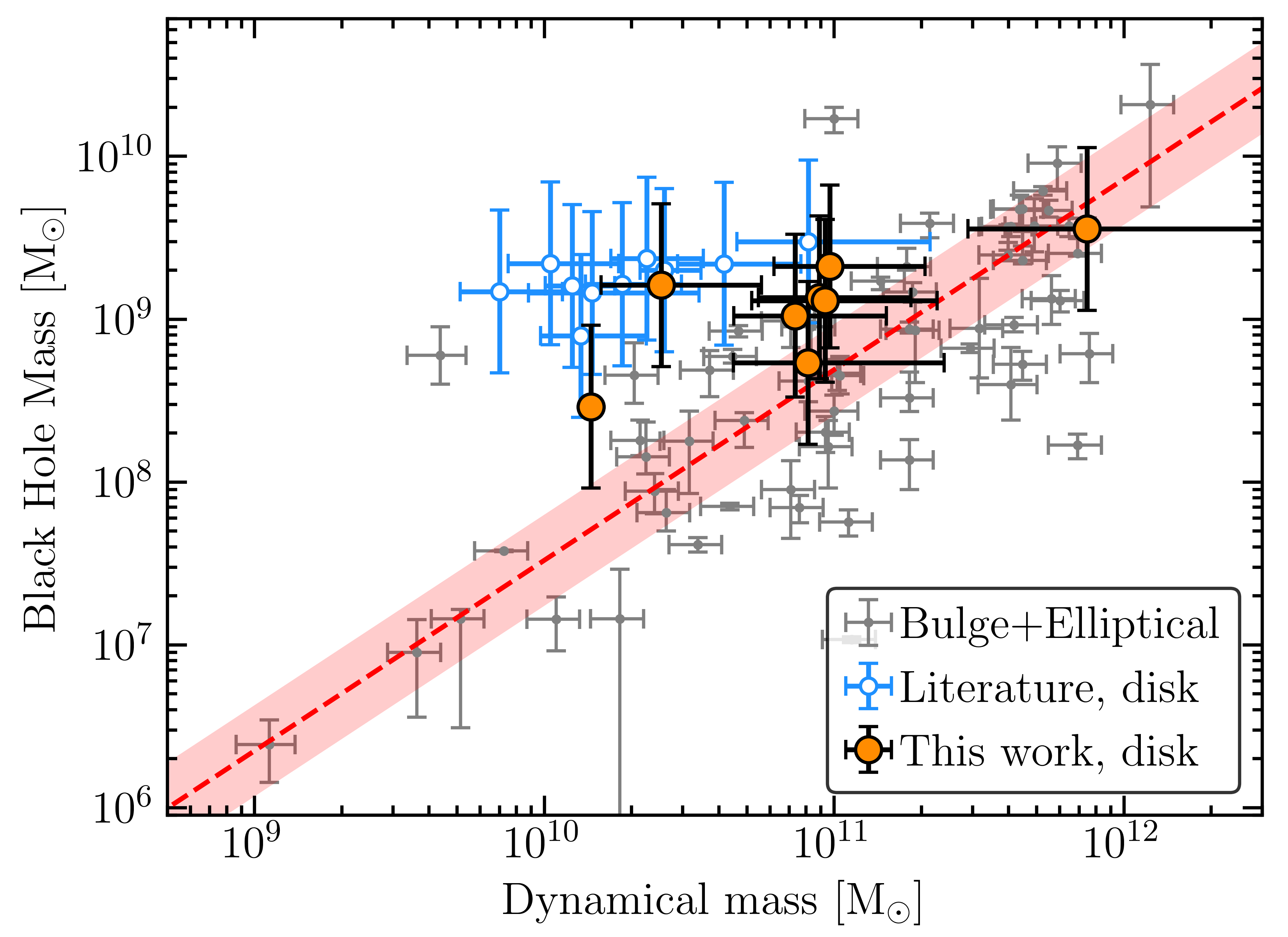}
\caption{
Same as the right panel of Figure \ref{fig:msigma}, but for the dynamical mass estimated from Equation \ref{eq:dynamical_N21}.
Most of the galaxies fall above the locally derived $M_{\rm BH}$--$M_{\rm bulge}$ correlation \citep[e.g.][]{KH13} with the black holes being over-massive by a factor of $\sim10$, although with substantial scatter. 
\label{fig:mdyn}}
\end{figure}

\subsection{SMBH--host Co-evolution} \label{subsec:co}
The recent measurement of the spatial density of the brightest quasars suggests an extraordinary rapid rise of the bright quasar population from $z\sim6.7$ to $z\sim6$, indicating a rapid buildup of SMBHs in a short time ($\sim$120 Myr) from $z>6.5$ to $z\sim6$ \citep{Wang19a}. Therefore, investigating the co-evolution between these earliest SMBHs and their host galaxies is crucial for understanding the assembly of early massive galaxies. Such investigation has been extensively explored in the past few years at $z\sim6$ \citep[e.g.][]{Wang13,Willott15,Izumi19,Neeleman21}, but limited to a small sample of quasars at $z>6.5$ \citep[e.g.][]{Venemans17a,Neeleman21}. In this section, we investigate the early co-evolution of SMBHs and their host galaxies with both the $M_{\rm BH}-\sigma$ and $M_{\rm BH}-M_{\rm dyn}$ relations by combining the sample of $z>6.5$ quasars studied in this work with data collected from the literature for $z\sim6$ quasars. 

The BH masses of all $z>6.5$ quasars studied in this work, except for J0213--0626, J0229--0808, and J0430--1445, are measured from the \ion{Mg}{2} emission line and are collected from the literature \citep{Banados21,Wang21b,Yang21}. We also collected the BH masses for other [\ion{C}{2}]-detected quasars where available \citep[e.g.][]{Onoue19,Shen19,Schindler20,Wang21b, Farina22}. 
To be consistent, all $M_{\rm BH}$ values are calculated using the single-epoch black hole mass estimator \citep{Vestergaard09} and under the same cosmological model used in this work. 

Since detecting the stellar light of these quasar host galaxies is challenging \citep[e.g.][]{Marshall20}, $\sigma$ and/or $M_{\rm dyn}$ are usually estimated from the [\ion{C}{2}] observations. 
First, we simply assume $\sigma={\rm FWHM}/(2~\sqrt{\rm 2~ln(2)})$ following most of the literature \citep[e.g.][]{Venemans17a}.
Secondly, \cite{Ho07} suggests that one can relate molecular or atomic gas in a disk to stellar bulges using the line width measured at 20\% of the peak intensity after correcting for the inclination and other effects. Following \cite{Willott15}, we set the [\ion{C}{2}] line full-width at 20\% equal to 1.5 times the FWHM since the lines are approximately Gaussian. We then estimate $\sigma$ following \cite{Ho07} by correcting the turbulent broadening and the inclination factor (assuming a median inclination angle of 45$^\circ$). 

In Figure \ref{fig:msigma}, we show the $M_{\rm BH}-\sigma$ and $M_{\rm BH}-M_{\rm dyn}$ relations for all $z>5.7$ quasars with both [\ion{C}{2}] detections and \ion{Mg}{2}-based black hole mass measurements as well as the measurements and derived relations from local galaxies \citep{KH13}. Overall the newly available measurements of $z>6.5$ quasars from this work occupy the same regions as those of $z\sim6$ quasars collected from the literature in all the plots.
In the left panel of Figure \ref{fig:msigma} we show the $\sigma$ derived from ${\rm FWHM}/(2~\sqrt{\rm 2~ln(2)})$ versus $M_{\rm BH}$. Similar to previous findings \citep[e.g.][]{Wang10,Wang16,Venemans17a, Wang19b}, most of the SMBHs are significantly over-massive compared to the local $M_{\rm BH}-\sigma$ relation. On the other hand, the deviation is smaller if we use the $\sigma$ derived following \cite{Ho07} as shown in the middle panel of Figure \ref{fig:msigma}. The reason is that the method proposed by \cite{Ho07} corrects for the inclination effect and thus has a larger $\sigma$ than the simple estimate from $\rm FWHM_{[CII]}$ \citep[see also][]{Wang10,Willott15}. Nevertheless, we caution that both methods have large uncertainties given the high turbulent velocities and unclear kinematics of the [\ion{C}{2}] emitting gas, and the large uncertainty on the inclination angle estimates. 
In the right panel of Figure \ref{fig:msigma}, we show the $M_{\rm BH}-M_{\rm dyn}$ relation, where $M_{\rm dyn}$ of rotation-dominated quasar host galaxies are estimated using Equation \ref{eq:dynamical}. The majority of $z\gtrsim6$ SMBHs are slightly over-massive compared with expectations from the local $M_{\rm BH}-M_{\rm bulge}$ relation, although some of them still follow the local trend. 

Again as noted by \cite{Neeleman21}, the dynamical mass estimated from Equation \ref{eq:dynamical} could be overestimated by a factor of $\sim2.3$. We show the $M_{\rm BH}-M_{\rm dyn}$ relation after correcting for this factor using Equation \ref{eq:dynamical_N21} in Figure \ref{fig:mdyn}. 
In Figure \ref{fig:mdyn}, we present the corrected $M_{\rm dyn}$ values for galaxies classified as rotation-dominated systems, encompassing both literature sources and our observations. Most of the quasars studied in this work, with the exception of J0910--0414, exhibit overmassive SMBHs when compared to local relations. The ratio $M_{\rm BH}/M_{\rm dyn}$ is approximately three times higher than the local relation on average, spanning a broad range of 1.2--14.5. The literature sample, comprising ten rotation-dominated quasar host galaxies, 
shows an even more pronounced BH mass excess on average (around 20 times higher). 
We observe that the disparities between these two samples arise from differences in the [\ion{C}{2}] emission sizes, as depicted in Figure \ref{fig:size}, and the slightly more rounded morphology of the quasar host galaxies studied in our work compared to those in the literature \citep[e.g.,][]{Venemans20,Neeleman21}. These differences can be attributed to the slightly higher resolution of literature observations, although with similar depth, leading to smaller [\ion{C}{2}] sizes due to missing extended emission and more complicated and resolved structures. Considering both samples jointly, the deviation of the $M_{\rm BH}-M_{\rm dyn}$ relationship for these high-redshift quasar host galaxies from local relations could be a factor of approximately 10 (indicating black holes being overmassive by a factor of 10), with a broad range spanning from approximately 0.6 to 60. We acknowledge that the current $M_{\rm dyn}$ measurements for both the literature sample and our work are subject to considerable uncertainties. Thus, we conclude that distant luminous quasars may indeed host overmassive black holes, but with substantial scatter and uncertainty in $M_{\rm dyn}$ measurements.
We further note that the $M_{\rm BH}-M_{\rm dyn}$ correlation could be affected by the Lauer bias \citep{Lauer07}, since overmassive objects are tend to be brighter and thus would be overrepresented in the selected sample \citep[see][for a detailed discussion]{Zhang23}.

\section{Summary} \label{sec:summary}
In this work, we present a spatially resolved [\ion{C}{2}] survey of 31 $z\sim7$ galaxies hosting luminous quasars with {ALMA} and {NOEMA}.  The purpose of this paper is to describe the program design and to present the data and analyses of [\ion{C}{2}] properties, continuum luminosity, star formation rate (SFR), the size and morphology of quasar host galaxies, as well as the first-order constraints on dynamical masses of the quasar host galaxies. We summarize our main findings below.

\begin{itemize}

\item We detect the [\ion{C}{2}] emission of 26 quasar host galaxies with a [\ion{C}{2}] luminosity range of $L_{\rm [CII]}=(0.3-5.5)\times10^9~L_\odot$, and detect the dust continuum of 27 quasar host galaxies with a far-infrared luminosity range of $L_{\rm FIR}=(0.5-13.0)\times10^{12}~L_\odot$. The SFR of these quasar host galaxies are in the range of 100--3000 $M_\odot~{\rm yr}^{-1}$ with a fiducial assumption of a graybody model with $T=47~K$ and $\beta=-1.6$.

\item Both $L_{\rm [CII]}$ and $L_{\rm FIR}$ are correlated ($\rho\simeq0.4$) with the quasar bolometric luminosity but with substantial scatter. 

\item The [\ion{C}{2}]/FIR luminosity ratio ranges from $\sim10^{-4}$ to $\sim10^{-2}$ and anti-correlates with the FIR luminosity ($L_{\rm FIR}$) or the FIR luminosity surface density ($\Sigma_{\rm FIR}$). This so-called `[\ion{C}{2}] deficit' is consistent with that found in lower redshift quasar host galaxies and infrared luminous galaxies.

\item The majority of our quasar host galaxies are clearly resolved with a median diameter of $\sim$5 kpc, and the [\ion{C}{2}] emission is usually less concentrated than the dust emission. 

\item The quasar host galaxies show diverse kinematics and morphologies, with $\sim40$\% of them showing a clear velocity gradient, consistent with a rotating gas disk, while the rest of them show either dispersion-dominated compact morphology or disturbed kinematics. 

\item Our first-order estimates of the dynamical masses of the rotation-dominated quasar host galaxies give $M_{\rm dyn}=(0.1-7.5)\times10^{11}~M_\odot$.
Considering our findings alongside those of literature studies, we conclude that the majority of black holes at the centers of distant luminous quasars are overmassive compared to the local $M_{\rm BH}-M_{\rm dyn}$ relation but with significant scatter, with the ratio between $M_{\rm BH}$ and $M_{\rm dyn}$ ranging from approximately 0.6 to 60 times that of local relations, and large uncertainties. 

\end{itemize}

In the next steps, we will perform kinematic modeling of [\ion{C}{2}] emissions to better constrain the dynamical mass of host galaxies. We will also stack the [\ion{C}{2}] emissions to detect the low surface brightness emissions and search for outflow signatures which has been suggested to be strong in these luminous quasars by the strong blueshift of high-ionization UV broad emission lines and the high broad absorption line quasar fraction \citep{Yang21}, as well as the presence of blueshifted [\ion{O}{3}] lines \citep{Yang23}. 
Furthermore, most of these quasars will be observed by ALMA mosaic observations from the ALMA Cycle 9 Large Program (2022.1.01077.L) which will enable us to characterize the Mpc-scale environment of these quasars. The majority of these quasars will also be observed by {\it JWST} which will allow us to detect the stellar light of these quasar host galaxies. 
Finally, combining the [\ion{C}{2}] systemic redshifts with the quasar optical-to-infrared spectroscopy \citep{Yang20b,Yang21} will allow us to constrain the cosmic reionization and quasar life time by measuring the quasar proximity zone sizes and modeling damping wing profiles.

\acknowledgments
We greatly thank the referee for their careful review and constructive comments. 
F.W and X.F. acknowledge support from the US NSF Grant AST-2308258. 
F.W. thanks the support provided by NASA through the NASA Hubble Fellowship grant \#HF2-51448 awarded by the Space Telescope Science Institute, which is operated by the Association of Universities for Research in Astronomy, Incorporated, under NASA contract NAS5-26555.
J. Y. and X. Fan acknowledges support by NSF grants AST 19-08284. 
E.P.F. is supported by the international Gemini Observatory, a program of NSF’s NOIRLab, which is managed by the Association of Universities for Research in Astronomy (AURA) under a cooperative agreement with the National Science Foundation, on behalf of the Gemini partnership of Argentina, Brazil, Canada, Chile, the Republic of Korea, and the United States of America. 
J.T.L. acknowledges the financial support from the National Science Foundation of China (NSFC) through the grant 12273111, and also the science research grants from the China Manned Space Project.
This paper makes use of the following ALMA data: ADS/JAO.ALMA\#2018.1.01188.S., ADS/JAO.ALMA \#2019.1.01025.S, ADS/JAO.ALMA\#2019.A.00017.S. ALMA is a partnership of ESO (representing its member states), NSF (USA) and NINS (Japan), together with NRC (Canada), MOST and ASIAA (Taiwan), and KASI (Republic of Korea), in cooperation with the Republic of Chile. The Joint ALMA Observatory is operated by ESO, AUI/NRAO and NAOJ. In addition, publications from NA authors must include the standard NRAO acknowledgement: The National Radio Astronomy Observatory is a facility of the National Science Foundation operated under cooperative agreement by Associated Universities, Inc.
This work is based on observations carried out under project number W18EJ with the IRAM NOEMA Interferometer. IRAM is supported by INSU/CNRS (France), MPG (Germany) and IGN (Spain).

\facilities{{\it ALMA}, {\it NOEMA}}

\software{
Astropy \citep{astropy},
CASA \citep{CASA},
GILDAS,
Matplotlib \citep[][\url{http://www.matplotlib.org}]{matplotlib},
Numpy \citep{numpy},
SciPy \citep{scipy}
}

\end{document}